\newcommand*{\ATLASLATEXPATH}{latex/}

\pdfoutput=1

\documentclass[UKenglish,texlive=2011,cernpreprint]{\ATLASLATEXPATH atlasdoc}

\usepackage[subfigure=true,biblatex=false]{\ATLASLATEXPATH atlaspackage}
\usepackage{dcolumn}
\usepackage{float}
\usepackage{cite}
\usepackage{\ATLASLATEXPATH atlascontribute}

\usepackage{\ATLASLATEXPATH atlasphysics}
\usepackage{\ATLASLATEXPATH atlasmisc}

\graphicspath{{logos/}{figures_local/}}

\usepackage{MultiPhotonPaper-defs}

\hypersetup{pdftitle={ATLAS draft},pdfauthor={The ATLAS Collaboration}}
\AtlasTitle{Search for new phenomena in events with at least three photons collected in $pp$ collisions at $\sqrt{s}$ = 8 TeV with the ATLAS detector}

\author{The ATLAS Collaboration}

\nolinenumbers

\AtlasRefCode{EXOT-2013-24}

\AtlasJournal{Eur.\ Phys.\ J.\ C.}
\PreprintIdNumber{CERN-PH-EP-2015-187}

\AtlasAbstract{%
Results of a search for new phenomena in events with at least three photons are reported.  Data from proton--proton collisions at a centre-of-mass energy of 8 TeV, corresponding to an integrated luminosity of 20.3 fb$^{-1}$, were collected with the ATLAS detector at the LHC.  The observed data are well described by the Standard Model.  Limits at the 95\% confidence level on new phenomena are presented based on the rate of events in an inclusive signal region and a restricted signal region targeting the rare decay $Z\to 3\gamma$, as well as di-photon and tri-photon resonance searches.  For a Standard Model-like Higgs boson with a mass of 125 GeV decaying to four photons via a pair of intermediate pseudoscalar particles ($a$), limits are found to be $\sigma \times {\text {BR}}(h \rightarrow aa) \times {\text {BR}}(a \rightarrow \gamma \gamma)^{2} < 10^{-3} \sigma_{\text{SM}}$ for 10 GeV $< m_{a} <$ 62 GeV.  Limits are also presented for Higgs boson-like scalars ($H$) for $m_{H} > $ 125 GeV, and for a $Z'$ decaying to three photons via $Z' \rightarrow a+\gamma \rightarrow 3\gamma$.  Additionally, the observed limit on the branching ratio of the $Z$ boson decay to three photons is found to be BR$(Z \rightarrow 3\gamma) < 2.2 \times 10^{-6}$, a result five times stronger than the previous result from LEP.
}

\begin{document}

\maketitle

\section{Introduction}
\label{sec:intro}

\begin{figure}[H]
  \centering
  \subfigure{
  \includegraphics[width=0.45\columnwidth]{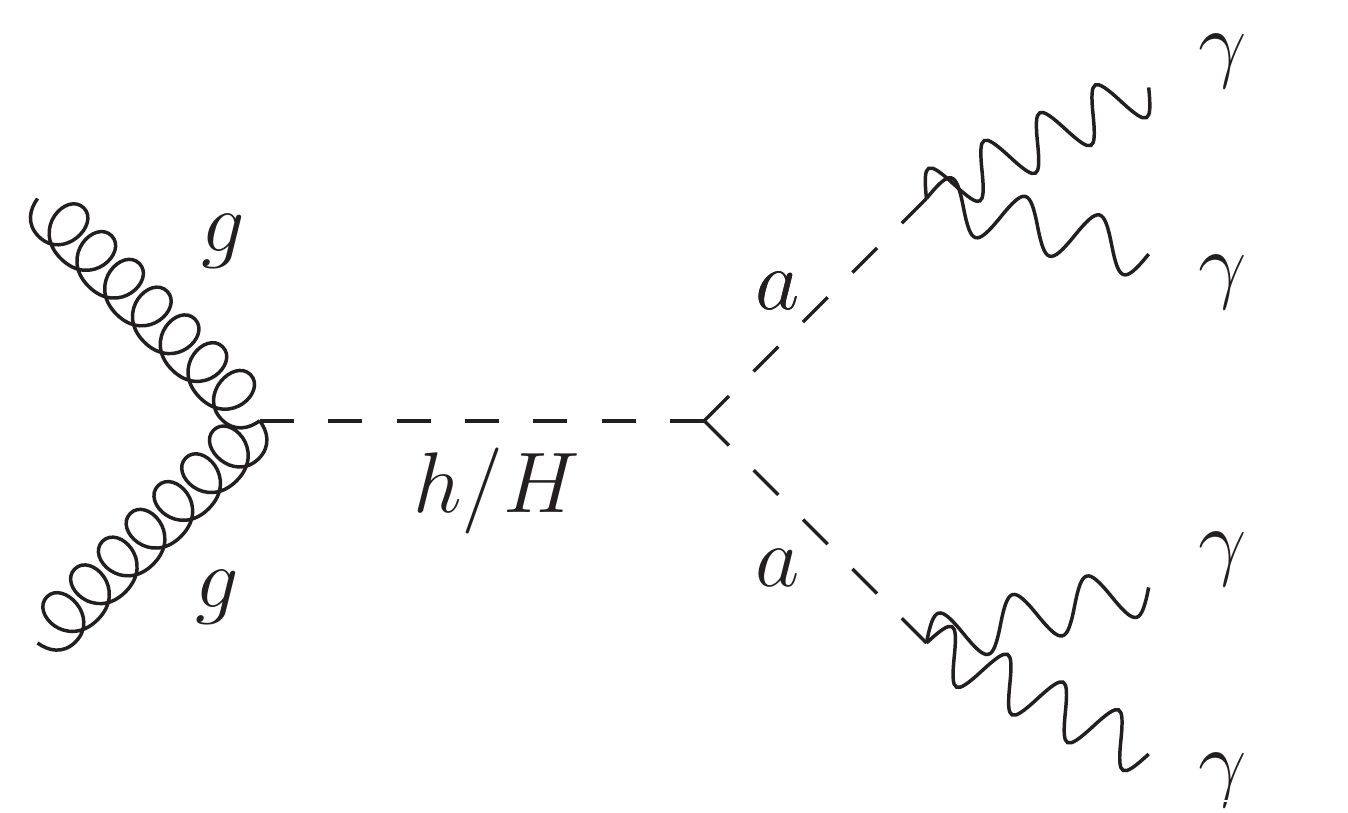}
  \label{fig:hToAATo4Ph}
  }
  \hskip 10mm
  \subfigure{
  \includegraphics[width=0.45\columnwidth]{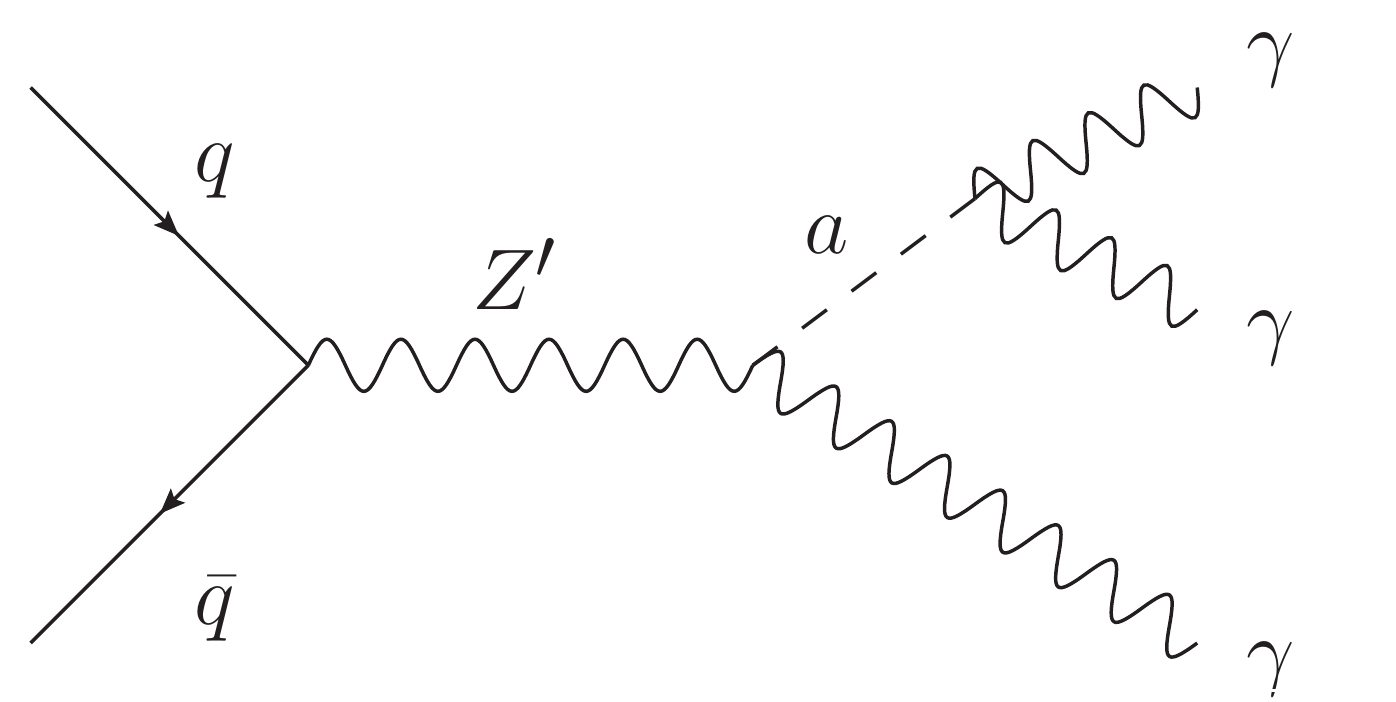}
  \label{fig:ZprimeTo3Ph}
  }
  \hskip 10mm
  \subfigure{
  \includegraphics[width=0.5\columnwidth]{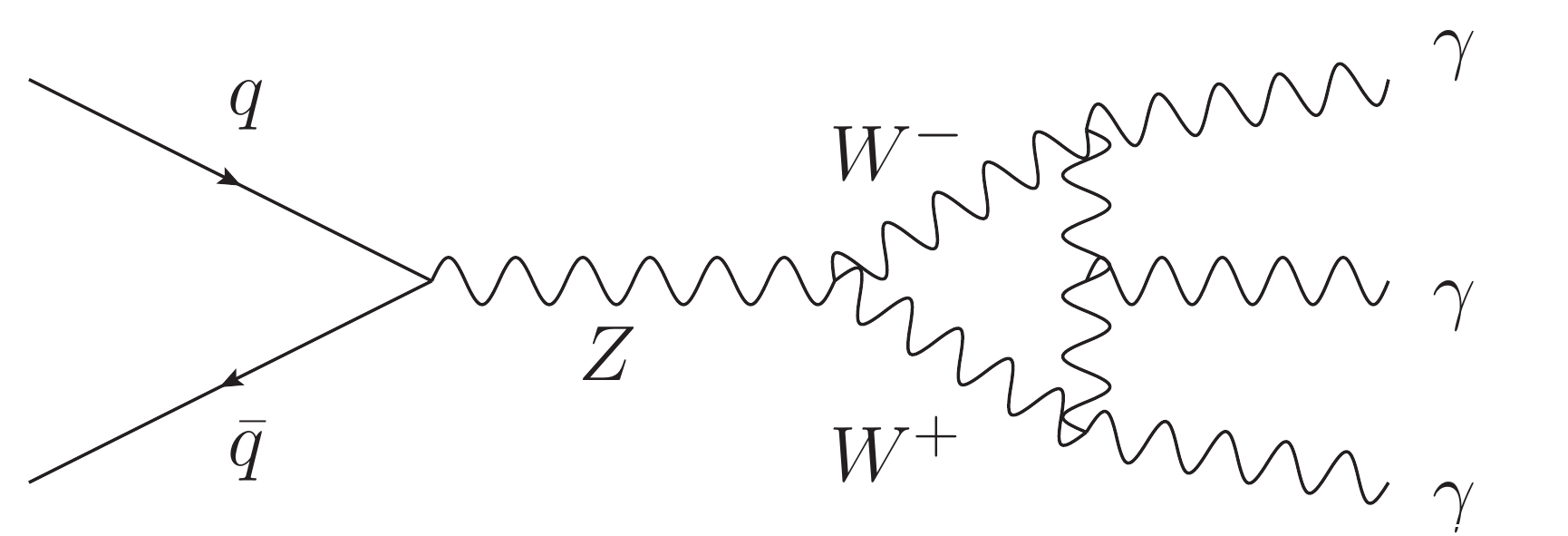}
  \label{fig:ZTo3Ph}
  }
  \caption{Feynman diagrams for possible beyond-the-Standard Model (top) and rare Standard Model (bottom) scenarios that result in final states with at least three photons.}
  \label{fig:BSMdiagrams}
\end{figure}

Many extensions of the Standard Model (SM) include phenomena that can result in final states consisting of three or more photons.  Extensions of the SM scalar sector~\cite{Chang:2006bw,Dobrescu:2000jt,Larios:2002ha,Curtin:2013fra,Dermisek:2007yt}, for example, often include pseudoscalar particles ($a$) with couplings to the Higgs boson~\cite{Aad:2012tfa,Chatrchyan:2012ufa} ($h$) and branching ratios into photons that would be visible at the LHC, in addition to scalars ($H$) with masses different from the SM-like Higgs boson of $m_{h} =$ 125 GeV that can also decay via $H \rightarrow aa \rightarrow 4\gamma$.  Other models feature additional vector gauge bosons that can decay to a photon and a new pseudoscalar boson, $a$, with the subsequent decay of the $a$ into a pair of photons, resulting in a three-photon final state~\cite{Toro:2012sv}.  Moreover, in the SM, the $Z$ boson can decay to three photons via a loop of $W^{\pm}$ bosons or fermions.  The decay is heavily suppressed and the branching ratio is predicted to be $\sim$5$\times$10$^{-10}$~\cite{Glover:1993nv}.  The current most stringent bound on this process comes from the L3 Collaboration, which placed a limit of BR$(Z \rightarrow 3\gamma) < 10^{-5}$~\cite{Acciarri:1994gb}.  The ATLAS detector has collected $\sim$10$^{9}$ $Z$ boson events, and thus an observation of this decay would indicate an enhancement of this decay rate and could be evidence of phenomena not predicted by the SM.  Feynman diagrams for some of these beyond-the-Standard Model (BSM) and rare SM scenarios are shown in Figure~\ref{fig:BSMdiagrams}.

To ensure sensitivity to these and other possible rare SM and BSM scenarios, an inclusive three-photon search is performed using 20.3 fb$^{-1}$ of LHC proton-proton collisions collected by the ATLAS detector in 2012 at a centre-of-mass energy of 8 TeV.  Such a model-independent search is the first of its kind, as are the interpretations for a Higgs boson decaying to four photons via two intermediate pseudoscalar $a$ particles (for a Higgs boson of $m_{h} =$ 125 GeV and for Higgs-like scalars of higher masses) and for three-photon resonances corresponding to a new vector gauge boson.

The dominant backgrounds include the irreducible component with three or more prompt photons, as well as the reducible components consisting of combinations of photons and electrons or hadronic jets misidentified as photons.  The contributions from events with jets which are misidentified as photons are calculated from data-driven methods, while simulation is used to estimate the contributions from the irreducible background and the reducible background originating from electroweak processes that lead to electrons which are misidentified as photons in the detector.  Collision data is used to derive corrections to the probability obtained from simulation that electrons are misidentified as photons.

\section{The ATLAS detector}
\label{sec:ATLASDetector}
\newcommand{\AtlasCoordFootnote}{%
ATLAS uses a right-handed coordinate system with its origin at the nominal interaction point (IP)
in the centre of the detector and the $z$-axis along the beam pipe.
The $x$-axis points from the IP to the centre of the LHC ring,
and the $y$-axis points upwards.
Cylindrical coordinates $(r,\phi)$ are used in the transverse plane, 
$\phi$ being the azimuthal angle around the beam pipe.
The pseudorapidity is defined in terms of the polar angle $\theta$ as $\eta = -\ln \tan(\theta/2)$.
}

The ATLAS experiment~\cite{atlas-detector} at the LHC is a multi-purpose particle detector
with a forward-backward symmetric cylindrical geometry and a near $4\pi$ coverage in 
solid angle.\footnote{\AtlasCoordFootnote}
It consists of an inner tracking detector surrounded by a thin superconducting solenoid
providing a \SI{2}{\tesla} axial magnetic field, electromagnetic (EM) and hadronic calorimeters, and a muon spectrometer.
The inner tracking detector covers the pseudorapidity range $|\eta| < 2.5$.
It consists of silicon pixel, silicon micro-strip, and transition radiation tracking detectors.
Lead/liquid-argon (LAr) sampling calorimeters provide EM energy measurements
with high granularity.
A hadronic (iron/scintillator-tile) calorimeter covers the central pseudorapidity range ($|\eta| < 1.7$).
The end-cap and forward regions are instrumented with LAr calorimeters
for both EM and hadronic energy measurements up to $|\eta| = 4.9$.
The muon spectrometer surrounds the calorimeters and is based on
three large air-core toroid superconducting magnets with eight coils each.
Its bending power ranges from \num{2.0} to \SI{7.5}{\tesla\metre}.
It includes a system of precision tracking chambers and fast detectors for triggering.
A three-level trigger system is used to select events.
The first-level trigger is implemented in hardware and uses a subset of the detector information
to reduce the accepted rate to at most \SI{75}{\kilo\hertz}.
This is followed by two software-based trigger levels that
together reduce the accepted event rate to \SI{400}{\hertz} on average
depending on the data-taking conditions during 2012.

\section{Event and object selection}
\label{sec:ObjAndEvtSel}

This search utilises a three-photon trigger that places a minimum requirement on the photon momentum in the plane transverse to the beam axis (transverse momentum, or \pT) of 15 GeV, applied on three photon candidates in the EM calorimeter.  Each candidate is additionally required to satisfy a set of {\it loose} photon identification criteria~\cite{ATLAS-CONF-2012-123}.  Stringent detector and data quality criteria are applied offline.  Events are required to contain at least one interaction vertex, with no additional vertex requirements.

Photon candidates must satisfy a pseudorapidity requirement of $|\eta|$ $<$ 2.37, excluding the transition region between the barrel and end-cap of 1.37 $< |\eta| <$ 1.52, and must satisfy requirements on the shape of the energy deposit in the calorimeter.  A photon candidate is rejected if the barycentre of its energy deposit is within a cone of $\Delta R \equiv \sqrt{(\Delta\eta)^{2} + (\Delta\phi)^{2}} <$ 0.15 around the barycentre of the energy deposit of a higher \pT photon candidate.  Finally, selected photon candidates are required to satisfy a more stringent set of identification criteria, known as {\it tight}~\cite{ATLAS-CONF-2012-123}.  Photon isolation is defined by the amount of transverse energy, \ETiso, deposited in the EM calorimeter within a cone of size $\Delta R$ around the photon candidate, excluding the energy of the photon candidate itself.  It is a powerful means of distinguishing between photons and hadronic jets misidentified as photons, since the energy clusters deposited by photons in the EM calorimeter tend to be narrower in the transverse direction than those deposited by jets.  Because minimum-bias proton--proton interactions in the same or nearby bunch crossings (pileup) can affect the calculated photon isolation energy, a correction is applied based on an event-by-event energy density pileup estimation.  This search uses an isolation cone of size $\Delta R <$ 0.4, and a correction to the \ETiso value of a photon candidate is made when another photon candidate passing the {\it tight} identification criteria is found within an annulus of 0.15 $< \Delta R <$ 0.4 around the photon candidate.  The correction consists of subtracting the \pT value of the other photon candidate found within the annulus from the \ETiso value of the photon candidate under consideration.  The final isolation criterion is $\ETiso < $ 4 GeV.

Events in the inclusive signal region are required to have at least three tightly identified and isolated photon candidates, where the two photon candidates with the highest transverse momentum must have $\pT >$ 22 GeV while the third highest must have $\pT >$ 17 GeV.  The restricted signal region, targeting the rare decay $Z \rightarrow 3\gamma$, is a subset of these events where an additional criterion of 80 GeV $< m_{3\gamma} <$ 100~GeV is placed on the invariant mass of the three-photon system.  The signal regions are supplemented by several control regions, where at least one of the photon candidates fails the isolation requirement.

\section{Simulated event samples}
\label{sec:simulation}

Simulated event samples are used to estimate several SM background processes in the search for excesses in the inclusive signal regions, as well as to model signal predictions for both the inclusive searches and the resonance searches for the specific BSM scenarios considered here.  The simulated BSM signal samples are also used to define a fiducial region for which the search criteria are largely model-independent.

\subsection{Simulated backgrounds}
The SM two-photon process is an irreducible background to a three-photon search because a third photon may arise from minimum-bias proton--proton interactions in the same bunch crossing.  The SM two-photon background is simulated with {\sc Pythia 8}~\cite{Sjostrand:2007gs}, and the three- and four-photon backgrounds are simulated with {\sc MadGraph 5}~\cite{Alwall:2011uj}, with {\sc Pythia 8} used for fragmentation and hadronisation.  The production of two, three and four photons in the SM contains large contributions from higher-order Feynman diagrams.  Thus, the two-, three- and four-photon simulated event samples calculated at leading order (LO) are multiplied by factors (``$K$-factors") determined from studies with generators that include next-to-leading order (NLO) contributions, namely {\sc MCFM 6.8}~\cite{Campbell:2014yka} and {\sc VBFNLO 2.7.0}~\cite{Baglio:2014uba,Arnold:2011wj,Arnold:2008rz}, using the parton distribution function (PDF) sets CTEQ6L1~\cite{Pumplin:2002vw} for the LO cross sections and CT10~\cite{PhysRevD.82.074024} and MSTW8NL~\cite{Martin:2009iq} for the NLO cross sections.  These $K$-factors are 1.9 $\pm$ 0.2 for the two-photon process and 3.3 $\pm$ 0.5 for the three-photon process.  The four-photon process is not included in NLO generators, and since the four-photon background is $\sim$10$^{-3}$ of the total background expectation in the inclusive signal region, the three-photon $K$-factor is applied to the four-photon background sample.  The uncertainties for these $K$-factors are determined by multiplying the renormalisation and factorisation scales independently by 2 and 0.5 and taking the largest deviations from the nominal value of the $K$-factor.

The reducible backgrounds where electrons are misidentified as photons originate from multiple sources.  Processes where a $Z$ boson decays to an $e^{+}e^{-}$ pair, accompanied by a photon not from the matrix element, are modelled with {\sc Powheg-Box 1.0}~\cite{Alioli:2008tz}, using {\sc Pythia 8} for fragmentation and hadronisation, and $Z$+$\gamma$ production is modelled with {\sc Sherpa 1.4.1}~\cite{Gleisberg:2008ta}.  Backgrounds from processes involving the leptonic decay of the $W$ boson in association with photons and/or hadronic jets are simulated with {\sc Alpgen}~\cite{Mangano:2002ea} and {\sc Herwig}~\cite{Corcella:2000bw,Corcella:2002jc}.  Possible mis-measurement of the rate of electrons misidentified as photons in simulation is addressed by comparing to electrons misidentified as photons from $Z \rightarrow e^{+}e^{-}$ events in data.

To obtain estimates of the rates at which true photons populate the regions of kinematic phase space assumed to be dominated by jets (used in the calculation of the systematic uncertainty for the data-driven estimate of jet backgrounds), a sample containing events with one hard-process quark or gluon and one prompt photon is simulated using {\sc Pythia} and the CTEQ6L1 PDF set.

The PDF sets for the simulated event samples of background processes used for the final background estimate in the inclusive search, for the {\sc MadGraph}, {\sc Pythia} and {\sc Alpgen + Herwig} samples, are taken from CTEQ6L1, while for the {\sc Powheg-Box} and {\sc Sherpa} samples the PDF sets are taken from CT10.

\subsection{Simulated signal processes}
The $Z\gamma\gamma\gamma$ effective vertex has been implemented with {\sc FeynRules}~\cite{Alloul:2013bka,Degrande:2011ua} and then used in a customised {\sc MadGraph 5} model which is employed to simulate events, using the CTEQ6L1 PDF set and {\sc Pythia 8} for fragmentation and hadronisation. Each of the two non-trivial, independent, lowest-order effective Lagrangians for this process~\cite{Stohr:1993uj} contains a dimensionful coupling constant, and the values of these constants have been calculated~\cite{Gutierrez-Rodriguez:2013eya} using the SM expected $Z \rightarrow 3\gamma$ branching ratio of 5.41$\times 10^{-10}$.  These SM values are used in the simulation.  The BSM process of a Higgs boson produced via gluon fusion and decaying to four photons via a pair of intermediate $a$ particles is simulated with {\sc Powheg-Box} and {\sc Pythia 8} (using the CT10 NLO PDF set).  The BSM process of a new vector gauge boson decaying to three photons via $Z' \rightarrow a+\gamma \rightarrow 3\gamma$ is simulated with {\sc Pythia 8} (using the MSTW2008LO~\cite{Martin:2009iq} leading-order PDF set).

\subsection{Minimum-bias interactions and the ATLAS detector simulation}
Minimum-bias proton--proton interactions in the same or nearby bunch crossings (pileup) are modelled with {\sc Pythia 8}, using the MSTW2008LO PDF set.  These pileup events are overlaid onto the hard-scattering process for all simulated signal and background samples to reproduce the distribution of the average number of interactions per bunch crossing observed over the course of data-taking in 2012.

All signal and background samples are processed with the full ATLAS detector simulation~\cite{Aad:1267853} based on {\sc Geant} 4~\cite{Agostinelli:2002hh} and reconstructed using the same software as that used for collision data.

\section{Background composition estimate}
\label{sec:backgrounds}

The backgrounds in the search for excesses in the inclusive signal regions are estimated from a combination of simulated samples (detailed in the previous section) and methods employing collision data.  The dominant backgrounds in the inclusive signal region are the irreducible SM two-, three- and four-photon processes, while for the $Z \rightarrow 3\gamma$ search channel, backgrounds involving electrons misidentified as photons are dominant.

\subsection{Backgrounds estimated from simulation}

The irreducible SM two-, three- and four-photon backgrounds, as well as backgrounds from processes involving electrons in the final state originating from $Z$ decays and those involving the leptonic decay of the $W$ boson in association with photons and/or hadronic jets, are estimated via simulation.  The third photon for the SM two-photon background process typically arises from pileup interactions, but can occasionally be a quark- or gluon-initiated jet radiated from the incoming partons which is misidentified as a photon.  Possible double-counting with the $2\gamma$ + 1--jet final state (estimated via a data-driven method described in the following section) is avoided by omitting from consideration events in the SM two-photon simulated sample where one of the three photon candidates is a jet, using generator-level information.  Possible mis-measurement of the rate of electrons misidentified as photons in simulation is addressed by comparing $Z \rightarrow e^{+}e^{-}$ processes in simulation and in data.  The per-electron scale factor is the ratio of the misidentification rate determined in data to that determined in simulated samples.  This scale factor is independent of \pT and $\eta$ for the ranges considered here, and is found to be 1.03 $\pm$ 0.04.

\subsection{Data-driven estimates of $2\gamma$ + 1--jet, $1\gamma$ + 2--jet, and 3--jet backgrounds}

A crucial aspect of the analysis is the data-driven estimate of the backgrounds where hadronic jets are misidentified as photons (hereafter called ``jet fakes"), i.e., SM processes that can produce $2\gamma$ + 1--jet, $1\gamma$ + 2--jet, and 3--jet events.  Collision data are used to derive efficiencies for photons passing the isolation criterion ($\epsilon_{\gamma}$) and rates at which jets are misidentified as isolated photons ($f_{\text{jet}}$).  These values of $\epsilon_{\gamma}$ and $f_{\text{jet}}$ are then used in a likelihood matrix method (described below) to estimate the jet backgrounds.

A sample of photon candidates consisting mainly of jet fakes is defined in the following way.  The standard {\it tight} and {\it loose} photon identification categories are augmented with a {\it medium} definition~\cite{Aad:2011mh}, intermediate between {\it tight} and {\it loose}.  The {\it medium} photon is defined by relaxing some EM shower shape requirements that provide high levels of rejection of jet fakes.  When the {\it medium} definition is combined with a further requirement that the photon candidates fail {\it tight} (the combination hereafter called {\it non-tight}), the result is a sample of photon candidates that is primarily composed of jet fakes.  This method presupposes that the \ETiso distribution of the {\it non-tight} sample is composed primarily of jet fakes, and that the subset of {\it tight} photons with higher values of \ETiso (the ``tail", here for $\ETiso > $ 7 GeV) is dominated by jet fakes.  Under these assumptions, the tail of the {\it non-tight} distribution is scaled to match that of the {\it tight} distribution, thus providing a determination of the contribution of jet fakes to the signal region, i.e., the collection of photons that pass both the {\it tight} and the isolation criteria.  The scaled {\it non-tight} distribution is then subtracted from the {\it tight} distribution.  Photon isolation efficiency, $\epsilon_{\gamma}$, is then calculated as the ratio of the number of isolated photons (those that satisfy $\ETiso <$ 4 GeV) to the total number of photons in the {\it tight} distribution after this subtraction has been performed.  The rate at which jet fakes are identified as photons, $f_{\text{jet}}$, is the ratio of the number of isolated photons to the total number of photons in the {\it non-tight} distribution.

The assumptions described above are validated using simulated samples of events containing photons and jets, described in Section~\ref{sec:simulation}.  Any collection of photon candidates consists of some combination of actual photons, which can be defined as ``true", and other objects that are misidentified as photons.  The non-zero true photon contamination in the set of {\it non-tight} photons, and the set of {\it tight} photons that fail the isolation criterion, is taken from the simulated samples and is used to derive a systematic uncertainty (described in Section~\ref{sec:systematics}) on the jet background estimate procedure.

The procedure is performed separately for three kinematic regions as follows.  Photons are ordered by \pT, highest to lowest.  Three regions in the $\pT$--$\eta$ plane are defined as 1) 15 GeV $< \pT <$ 40 GeV and $|\eta| < 1.37$, 2) $\pT > $ 40 GeV and $|\eta| < 1.37$, and 3) $1.52 < |\eta| < 2.37$.  The separation into lower and higher \pT bins around 40 GeV is chosen because this is the value at which $\epsilon_{\gamma}$ and $f_{\text{jet}}$ are changing rapidly, and the three regions were chosen to maintain a large number of events in each bin.  The values of $\epsilon_{\gamma}$ and $f_{\text{jet}}$ are then calculated for each of the three regions, and the results are shown in Table~\ref{tab:EffFRABCTopo}.

\begin{table}
  \begin{center}
    \begin{tabular}{ccc}
      \hline \hline
	Kinematic region &  \multicolumn{2}{c}{Fraction satisfying isolation criterion} \\
	\cmidrule(lr){2-3}
	 &  \multicolumn{1}{c}{Photons ($\epsilon_{\gamma}$)}  &  \multicolumn{1}{c}{Jets misidentified as photons ($f_{\text{jet}}$)} \\
	\hline
	1: 15 GeV $< \pT < $ 40 GeV, $|\eta| <$ 1.37	&  0.939 $\pm$ 0.007 $\pm$ 0.009  &  0.424 $\pm$ 0.001 $\pm$ 0.013 \\
	2: $\pT > $ 40 GeV, $|\eta| <$ 1.37  		&  0.906 $\pm$ 0.006 $\pm$ 0.013  &  0.256 $\pm$ 0.002 $\pm$ 0.010 \\
	3: 1.52 $< |\eta| <$ 2.37  				&  0.933 $\pm$ 0.007 $\pm$ 0.009  &  0.431 $\pm$ 0.002 $\pm$ 0.013 \\

      \hline \hline
    \end{tabular}
  \caption{Photon isolation efficiencies ($\epsilon_{\gamma}$) and rates of jets misidentified as photons ($f_{\text{jet}}$) from collision data for the three kinematic regions, used for the jet background estimate.  The isolation criterion is $\ETiso <$ 4 GeV.  The three regions were chosen to maintain a large number of events in each bin.  The first uncertainty is statistical while the second is systematic.}
  \label{tab:EffFRABCTopo}
  \end{center}
\end{table}

The data-derived $\epsilon_{\gamma}$ and $f_{\text{jet}}$ values are applied to events with three photon candidates to estimate the SM $2\gamma$ + 1--jet, $1\gamma$ + 2--jet, and 3--jet backgrounds.  This is done using a likelihood-based version of a standard matrix method (here called the ``likelihood matrix method").  In standard matrix methods~\cite{Aad:2011mh}, a matrix of efficiencies relates an observed event that falls into a particular event category (based on some discriminating variable or variables) to the true, unknown final states to which the event has the possibility of corresponding, and the matrix is inverted to determine probabilities that a given observed event corresponds to one of these true final states.  When summed over a large number of events, these per-event estimators average to the overall estimate of the number of events in each true final state.

In the likelihood matrix method, by contrast (and with respect to the present three-photon search), the expected yield for each three-object final state consisting of jets plus photons or all jets is the result of fitting a likelihood function to data.  For the event sample where all three photons, ordered from highest to lowest $\pT$, have satisfied the {\it tight} requirements, events are placed into 160 orthogonal categories designated by six criteria.  These are defined first by the three regions in the $\pT$--$\eta$ plane to which each photon candidate belongs.  These are the same three regions that are used to categorise photons and to calculate $\epsilon_{\gamma}$ and $f_{\text{jet}}$, described and labeled previously as regions 1, 2, and 3.  The remaining three criteria by which each event is categorized are three boolean variables, one for each photon candidate, indicating whether it passed or failed the isolation criterion.  Since each of the three photon candidates either passes (P) or fails (F) isolation, there are $2^{3} = 8$ possible isolation combinations for three photons: PPP, PPF, PFP, FPP, PFF, FPF, FFP, and FFF.  The three photons are ordered by $\pT$, from highest to lowest, and, since one of the three kinematic regions defined above depends only upon $\eta$, there are twenty possible $\pT$--$\eta$ bin combinations: 333, 332, 323, 322, 331, 313, 311, 321, 222, 223, 232, 233, 221, 211, 231, 213, 111, 113, 131, 133.   This results in 8$\times$20 = 160 categories, denoted PPP\_333 for those events where the three photon candidates all passed isolation and had $\pT$--$\eta$ values placing them in the ``3" kinematic region, PPF\_321 for those events where the leading and subleading photons passed isolation and the sub-subleading photon failed isolation, and the $\pT$--$\eta$ value combinations placed them successively in the ``3", ``2", and ``1" regions, etc.

Each of the 160 categories corresponds to a Poisson function where the observed number of events is the number of events seen in data for that category and the expected number of events is a sum of terms corresponding to each of the possible true (unknown) final states consisting of photons and jets or only jets for a particular $\pT$--$\eta$ combination.  Each term in a given sum is multiplied by the appropriate values of $\epsilon_{\gamma}$ and $f_{\text{jet}}$.  A likelihood is then constructed consisting of a product of the 160 Poisson functions.  The expectations for each true final state are the maximum likelihood estimators that result from fitting this likelihood function to the data.  That is, the true unknown expectations are allowed to float in the fit and are constrained to be positive and, hence, physical.  The estimated number of events of a given final state in a particular signal or control region --- defined by whether the photons passed or failed isolation --- is determined by summing the resulting expectations from the fit times the appropriate $\epsilon_{\gamma}$ and $f_{\text{jet}}$ values.

\section{Systematic uncertainties}
\label{sec:systematics}

\subsection{Data-driven uncertainties}

For the data-driven jet background estimate, systematic uncertainties arise in the calculation of the rate of photons passing the isolation criterion, $\epsilon_{\gamma}$, and the rate of jets misidentified as isolated photons (``jet fakes"), $f_{\text{jet}}$.  This calculation relies upon the assumptions that both the tail ($\ETiso > $ 7 GeV) of the $\ETiso$ distribution of {\it tight} photons and the entirety of the $\ETiso$ distribution of {\it non-tight} photons are primarily composed of jet fakes.  Tests on simulations of photons and jets indicate that the true photon contamination in these jet-dominated regions is between 5\% and 15\%, depending on the region.  These values are used to calculate different values of $\epsilon_{\gamma}$ and $f_{\text{jet}}$ (where the number of photon candidates in a given region is altered by the corresponding percentage) which are then used in the jet background estimate.  The deviations from the nominal signal region yield (assuming no true photon contamination) are calculated separately for the three final states of $2\gamma$ + 1 jet, $1\gamma$ + 2 jets, and 3 jets, and these values (4\%, 10\% and 21\%, respectively) are taken as systematic uncertainties on the estimates of these backgrounds in the signal region.

An additional uncertainty associated with the data-driven methods employed arises from the choice of kinematic variable used to categorise photons and then to calculate and apply $\epsilon_{\gamma}$ and $f_{\text{jet}}$.  The baseline analysis uses three bins in the $\pT$--$\eta$ plane, described in Section~\ref{sec:backgrounds}, and separate analyses using either \pT-dependence only or $\eta$-dependence only  are conducted as well.  The largest deviation of the two different methods from the nominal method is 13\%, which is taken as a systematic uncertainty.

\subsection{Simulation uncertainties}

The uncertainty on the integrated luminosity for the data sample is 2.8\%, derived using the same methodology as that
detailed in Ref~\cite{Aad:2013ucp}.

The photon identification efficiency has been directly measured in data using photons from $Z \rightarrow e^{+}e^{-}/\mu^{+}\mu^{-}$ radiative decays~\cite{ATLAS-CONF-2012-123}.  The systematic uncertainties on the signal region yield due to the uncertainty on this efficiency measurement are found to range from $<$1\% to 6\% for simulated backgrounds, and from 3\% to 7\% for simulated signal processes, depending on the sample.

As mentioned in Section~\ref{sec:ObjAndEvtSel}, the analysis supplements the isolation prescription --- $\ETiso <$ 4 GeV, with a cone size of $\Delta R <$ 0.4 --- with an isolation energy correction that is applied to photons with overlapping isolation cones.  This procedure improves sensitivity to lower-mass two-photon resonances where the photon pairs are close together in $\Delta R$.  To account for possible over- or under-correction due to a photon being near the edge of the isolation annulus, an additional systematic uncertainty is assessed.  The \pT values of all isolated photons in simulated samples are calibrated to yield agreement with the values observed in data~\cite{Aad:2014nim}.  Since the calibration factors for isolated photons deviate from one by typically less than 5\%, a value of 5\% is a conservative estimate of the uncertainty on photon \pT.  To assess the systematic uncertainty on the isolation energy correction, the measured value of the \pT of the other {\it tight} photon in the isolation cone is varied by 5\%, the correction procedure is applied, and the effects are propagated to the final event selection in the signal region.  For example, using simulated samples of a Higgs boson decaying to four photons via $H \rightarrow aa \rightarrow 4\gamma$, the systematic uncertainty due to this effect is smaller for higher ratios of $m_{a}/m_{H}$ (as large as 6\% when the $\pT$ is varied by $-5\%$ and $<$1\% when the \pT is varied by $+5\%$, for $m_{H}$ = 900 GeV and $m_{a}$ = 440 GeV), and the uncertainty is larger for smaller ratios of $m_{a}/m_{H}$ (as large as 69\% when the \pT is varied by $-5\%$ and 12\% when the \pT is varied by $+5\%$, for $m_{H}$ = 900 GeV and $m_{a}$ = 50 GeV), as the photons tend to overlap within the isolation cone more frequently.

The uncertainties on the event yields due to systematic uncertainties in the photon energy scale and resolution~\cite{Aad:2014nim} are found to range from $<$1\% to 4\% for the simulated signal and background samples.  The uncertainties on the event yields due to systematic uncertainties in the scale factors used to yield agreement between photon identification efficiencies calculated in data and simulated samples~\cite{Aad:2014nim} are found to range from $<$1\% to 8\% for simulated backgrounds, and from 1\% to 4\% for simulated signal processes.  The systematic uncertainty on the scaling factor for electrons misidentified as photons in simulated samples is taken to be the statistical uncertainty arising from the calculation, i.e., 4\%, since this is as large as or larger than the systematic uncertainties due to the photon energy scale and resolution, above.  The efficiency and uncertainties of the three-photon trigger chain have been determined to be 98.5\% $\pm$ 0.1\%~(stat.) $\pm$ 0.2\%~(syst.).  The trigger efficiency is calculated using single photons (with \pT values corresponding to the values used for the analysis event selection) from $Z$ boson radiative decays and then, under the assumption that the per-event performance of the photon trigger for one photon is uncorrelated to that for another photon in the same event, multiplying these values to obtain the overall trigger efficiency.

Uncertainties on calculated cross sections for simulated background processes due to QCD renormalisation and factorisation scales and due to the choice of PDF set and value of $\alpha_{\text s}$ used in simulation are addressed via the recommendations of {\sc PDF4LHC}~\cite{Botje:2011sn}.  The resulting combined uncertainties are found to range from 7\% to 16\%, depending on the simulated background process.  The total theoretical uncertainty on the SM 3$\gamma$ background process due to the uncertainty on the LO to NLO correction, combined with the uncertainties due to choice of PDF set and renormalisation and factorisation scales, is found to be $\sim$30\%.

Uncertainties exist for the measured or calculated production cross sections for SM particles for which BSM decays are considered as signal scenarios and are accounted for.  For the BSM Higgs boson scenario of $h \rightarrow aa \rightarrow 4\gamma$ the gluon fusion production cross section for the SM Higgs boson with $m_{h} =$ 125~GeV~\cite{Dittmaier:2011ti,Dittmaier:2012vm}, $\sigma_{h,\text{SM}} =$ 19.27 pb is used, with an uncertainty of $\pm 10.4$\% due to choice of PDF set and renormalisation and factorisation scales.  For the rare decay $Z \rightarrow 3\gamma$, the measured $pp \rightarrow Z$ cross section of $(2.79 \pm 0.02 \pm 0.11)\times 10^{4}$ pb~\cite{Aad:2011dm} is used.  An additional uncertainty of $\pm 12.3$\% --- determined by varying the QCD renormalisation and factorisation scales, PDF set, and value of $\alpha_{\text s}$ --- is also assessed, to account for variations in the simulation of the kinematics of the final-state photons and, hence, the acceptance in the signal region.

These systematic uncertainties are summarized in Table~\ref{tab:Systs}.

\begin{table}
  \begin{center}
    \begin{tabular}{ccc}
      \hline \hline

Data-driven       					& Background 	& Signal \\
\hline
Photon contamination of control regions 	& 4--21	& --- \\
Kinematic parametrization 			& 13 			& --- \\
\hline
Simulation 						& 			& \\
\hline
Photon ID 						& 1--6	 	& 3--7 \\
Photon isolation correction 			& $<$1--4 	& $<$1--69 \\
Photon energy scale and resolution 		& $<$1--4 	& $<$1--4 \\
Photon scale factors 					& $<$1--8 	& 1--4 \\
Electron scale factors 				& 4 			& 4 \\
Trigger 							& 0.2 		& 0.2 \\
Luminosity 						& 2.8 		& 2.8 \\
Cross section 						& 7--16 		& 4--12 \\
\hline

      \hline \hline
    \end{tabular}
  \caption{Systematic uncertainties (\%) on the expected event yields in the signal region.  The values given for data-driven backgrounds correspond to the three jet backgrounds described in the text.  For simulated samples, when a range is given it corresponds to the smallest and largest uncertainties for all simulated backgrounds or signals.}
  \label{tab:Systs}
  \end{center}
\end{table}

\section{Results and interpretations}
\label{sec:result}

This section presents results based on the number of events in a broad inclusive region and a restricted region focusing on the rare decay $Z\to 3\gamma$, as well as results from the search for resonances in the di-photon and tri-photon invariant mass spectra.

\subsection{Inclusive and $Z\to 3\gamma$ regions}
\begin{figure}
  \centering
  \subfigure{
  \includegraphics[width=0.45\columnwidth]{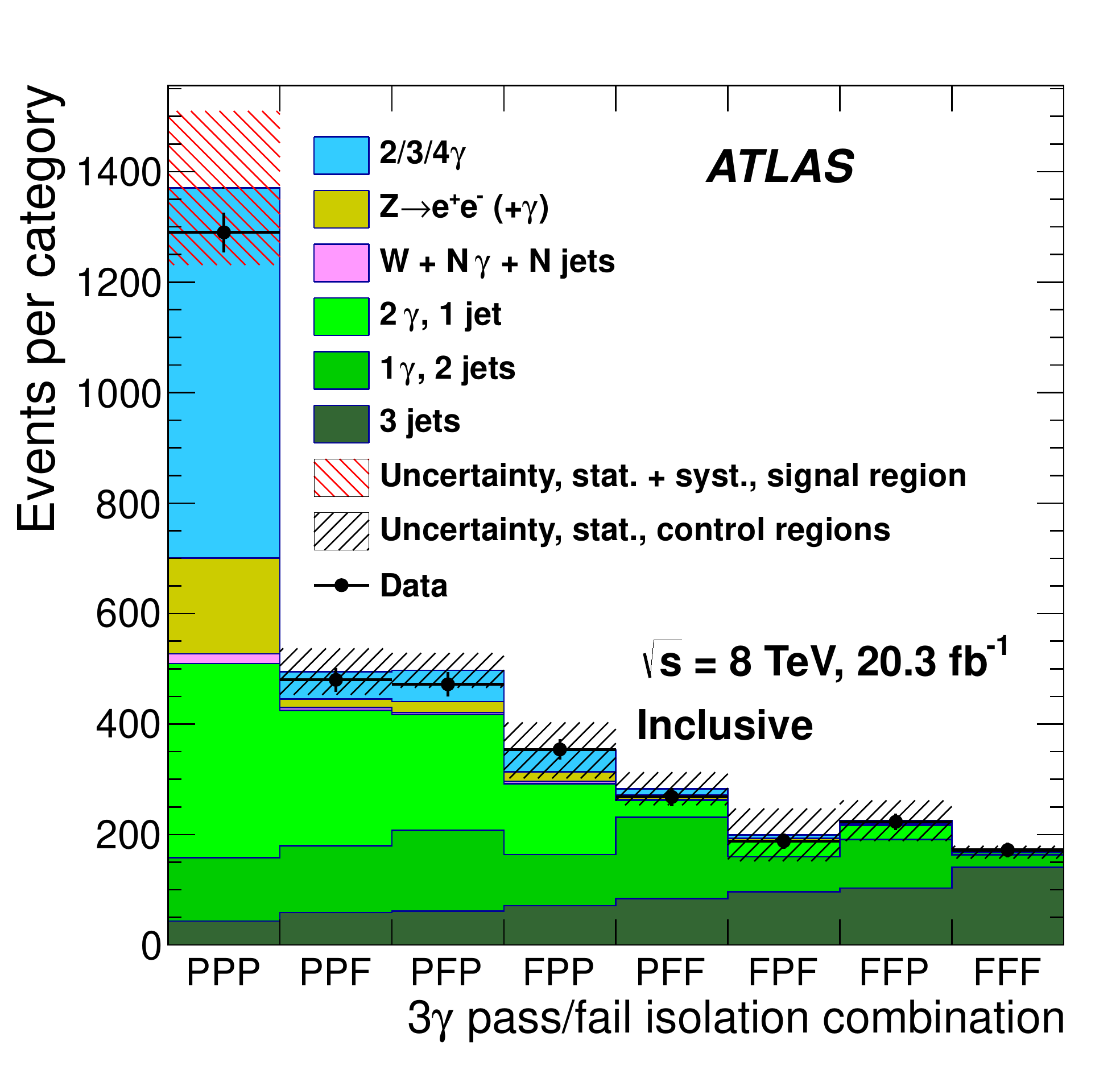}
  }
  \hskip 10mm
  \subfigure{
  \includegraphics[width=0.45\columnwidth]{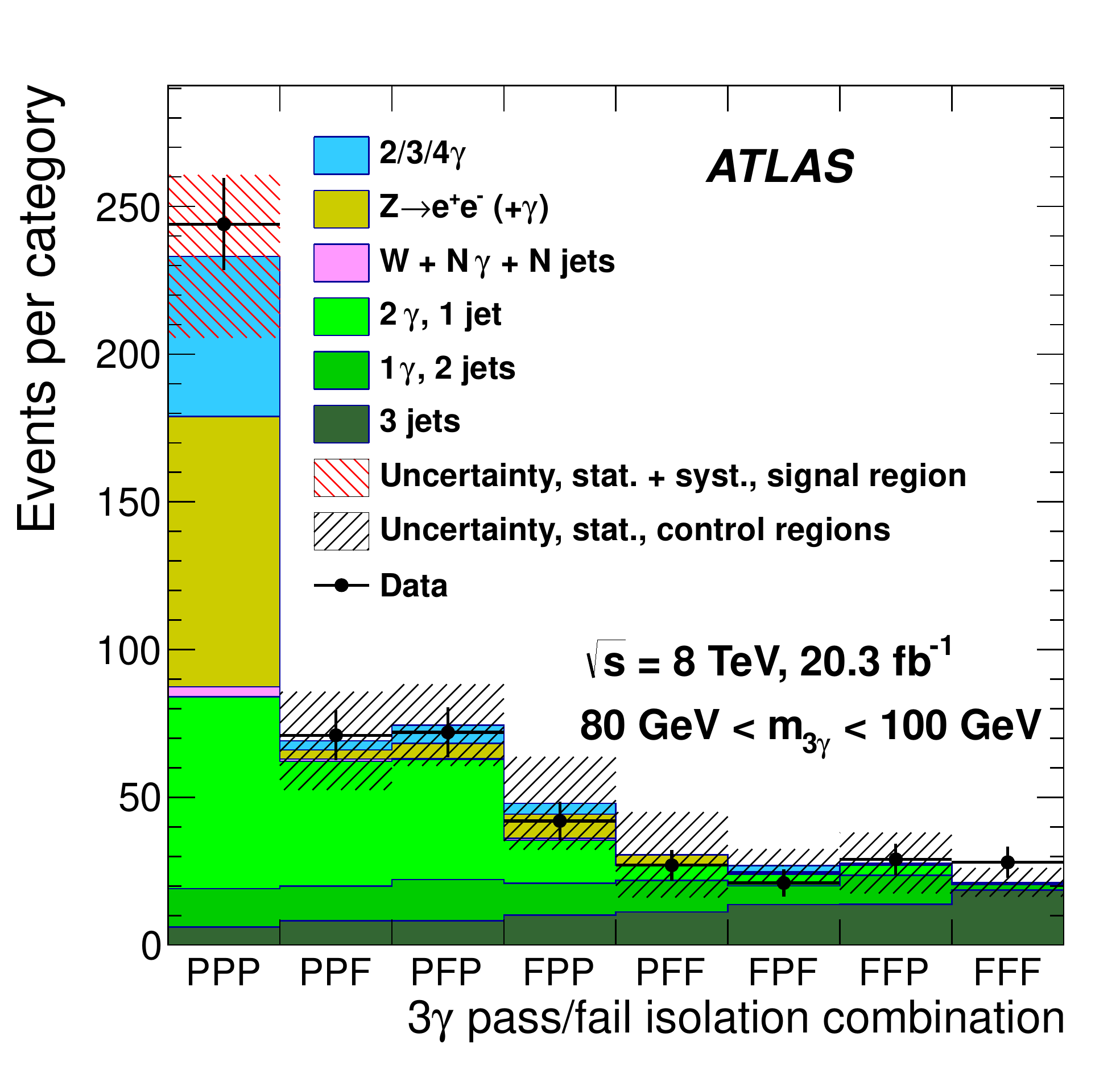}
  }
    \caption{Observed and expected yields in signal and control regions for the full mass range (left) and the restricted range of 80 GeV $< m_{3\gamma} <$ 100 GeV (right), for events where all three photon candidates satisfy the {\it tight} photon identification criteria.  The bins along the horizontal axis correspond to orthogonal subsets of events where each subset is categorised by whether the three photons --- ordered from largest to smallest values of \pT --- passed (``P") or failed (``F") the isolation criterion.  The leftmost bin is the signal region, composed of events satisfying PPP, and the other bins are the different control regions, where at least one of the photon candidates failed the isolation criterion.  The red hatched band, in the signal region bin, is the combination of statistical and systematic uncertainties, while the black hatched bands represent statistical uncertainties.  As a result of the data-driven jet background estimate, the statistical uncertainty in each bin is partially correlated with the uncertainty on the data in that bin.}
    \label{fig:ResultsTTT}
\end{figure}

The number of SM background events expected in the signal region is 1370 $\pm$ 140 (combined statistical and systematic uncertainties) and the observed number of events is 1290.  The observation is in agreement with the SM expectation.  Additionally, while the event selection is optimised for a search for physics beyond the SM as opposed to a measurement of the $3\gamma$ inclusive cross section, the results are nevertheless compatible with the irreducible all-photon process expectations from the SM.  The expected and observed yields in the signal region are presented in Table~\ref{tab:ResultsExpObs}, and the expected and observed yields in signal and control regions where all three photons have passed the {\it tight} identification criterion are shown in Figure~\ref{fig:ResultsTTT}.  In the figure, the red hatched band, in the signal region bin, is the combination of statistical and systematic uncertainties on all background sources, while the black hatched band, in the control regions, is the combination of statistical uncertainties of the data-driven jet background estimate and the expected yields from simulated samples of SM background processes.  For the inclusive signal region, this corresponds to a model-independent observed (expected $\pm 1\sigma$) upper limit, at the 95\% confidence level (C.L.), on the number of signal events of 240 (273$^{+83}_{-66}$), and to the model-dependent upper limits on the inclusive fiducial cross section in the aforementioned acceptance for the signal scenarios of the BSM Higgs boson and Higgs boson-like decays and the $Z'$ decays shown in Table~\ref{tab:ResultsFidSR}, where hypothesis testing and limit setting are calculated using the profile likelihood ratio as the test statistic for the $CL_{s}$ technique~\cite{0954-3899-28-10-313} in the asymptotic approximation~\cite{Cowan:2010js,Cowan:2013EPJC}.  The fiducial efficiencies for each signal scenario are determined with respect to a generator-level kinematic region with the same requirements applied to three-photon events as those used for the analysis signal region.  This fiducial region is defined as the set of events that contain three photons where 1) each photon satisfies a pseudorapidity requirement of $|\eta|$ $<$ 2.37, excluding the transition region between the barrel and end-cap of 1.37 $< |\eta| <$ 1.52, 2) the three photons satisfy $\pT >$ 22, 22, and 17 GeV, and 3) each photon satisfies $\ETtruthiso <$ 4 GeV, where \ETtruthiso is a generator-level definition of the photon isolation criterion equivalent to that used for event selection on reconstructed events.  The fiducial efficiencies are similar for the considered signal scenarios for mass points where the distributions of photon \pT, for all photons, tend to peak higher than $\pT >$ 50 GeV.  This is because the overall photon identification efficiency decreases for photons with $\pT <$ 50 GeV~\cite{ATLAS-CONF-2012-123}.  Since the \pT distribution for at least one of the photons for signal scenarios with lower-mass resonances tends to peak at lower values, the fiducial efficiencies are lower.

\newcolumntype{d}[1]{ D{.}{.}{-1} }
\newcolumntype{q}[1]{ D{.}{.}{3.1} }

\newcolumntype{L}[1]{ >{\raggedleft\arraybackslash}D{.}{.}{3.1} }

\begin{table}[h]
	\centering
	\sisetup{ table-number-alignment=center,
	table-alignment=center,
	separate-uncertainty=true, group-digits=false
	}

    \begin{tabular}{ l S[table-format=4.1,table-alignment=center]@{$\,\pm \,$} S[separate-uncertainty,table-figures-uncertainty=1,table-alignment=center] S[table-format=3.2,table-alignment=center]@{$\, \pm \,$} S[separate-uncertainty,table-figures-uncertainty=1,table-alignment=center] }

      	\hline \hline
      	\multicolumn{1}{l}{Process} &\multicolumn{2}{l}{Inclusive signal region} &\multicolumn{2}{l}{80 GeV $< m_{3\gamma} <$ 100 GeV} \\ 

	\hline

2$\gamma$ (sim.) 				&  330 	& 50  	& 24 		& 8  		\\
3$\gamma$ (sim.) 				&  340 	& 110  	& 30 		& 10  	\\
4$\gamma$ (sim.) 				&  1.3 	& 0.4 	& 0.07 	& 0.02 	\\
2$\gamma$,1j (D-D) 				&  350 	& 60 		& 65 		& 19 		\\
1$\gamma$,2j (D-D) 				&  110 	& 40 		& 13 		& 10 		\\
3j (D-D) 						&  43 	& 11 		& 6.1 	& 2.0 	\\
$Z\rightarrow e^{+}e^{-}$ (sim.)		&  85 	& 22 		& 43 		& 13 		\\
$Z$+$\gamma$ (sim.) 			&  89 	& 11 		& 48 		& 6 		\\
$W$+$\gamma$+(0,1,2)j (sim.) 	&  11.4 	& 1.5 	& 2.7 	& 0.7 	\\
$W$+2$\gamma$+(0,1,2)j (sim.) 	&  6.1 	& 0.5 	& 0.68 	& 0.08 	\\
\hline
Total SM exp. 					&  1370 	& 140 	& 233 	& 28 \\
\hline

      	\multicolumn{1}{l}{Observed} &\multicolumn{2}{l}{1290} &\multicolumn{2}{l}{244} \\ 
	\hline \hline
      	\end{tabular}
      
  \caption{Expected and observed event yields in the inclusive signal region and for the signal region with a further requirement of 80 GeV $< m_{3\gamma} <$ 100 GeV.  Background expectations estimated via simulations are
marked sim., whereas data-driven calculations are denoted as D-D.  The uncertainties for each row are the combination of statistical and systematic uncertainties for a given background process, and the overall uncertainties in the second to last row are the combined uncertainties for the total background expectations for each signal region.}
  \label{tab:ResultsExpObs}

\end{table}

\begin{table}
  \begin{center}
  \sisetup{retain-explicit-plus}
  \sisetup{round-mode = figures,round-precision=2,retain-explicit-plus=true}
    \begin{tabular}{ c c c c c }
     \hline \hline
      	\multicolumn{2}{c}{Expected background}    &\multicolumn{1}{c}{Observed} 	&\multicolumn{2}{c}{Obs. (exp.) 95\% C.L. upper limit on $N_{\text{sig}}$} \\ 

	\hline
      	\multicolumn{2}{c}{1370 $\pm$ 140}    &\multicolumn{1}{c}{1290}  	 	&\multicolumn{2}{c}{240 $\left(273\substack{+83 \\ -66}\right)$} \\ 
     	\hline \hline
      	\multicolumn{2}{l}{Signal process}    	&\multicolumn{1}{c}{Fiducial} 	&\multicolumn{1}{c}{Obs. (exp.) upper limit,} 			&\multicolumn{1}{c}{Obs. (exp.) upper limit,} \\ 
	\multicolumn{2}{l}{}  	&\multicolumn{1}{c}{efficiency} 	&\multicolumn{1}{c}{$\sigma_{\text{fid}} \times \mathcal{A}$ [fb]} 	&\multicolumn{1}{c}{$\sigma_{\text{overall}}$ [fb]} \\ 
	\hline       
	\multicolumn{2}{l}{$h/H \rightarrow aa \rightarrow 4\gamma$} & \multicolumn{1}{c}{} & \multicolumn{1}{c}{} & \multicolumn{1}{c}{} \\ 

$m_{h/H}$ [GeV] & $m_{a}$ [GeV] &    &    & \\ 
\hline
125 & 10     & $ 0.374\pm0.005 $ & {\num[round-precision=2]{31.611}} $\left(\,{\num[round-precision=2]{35.937}}\substack{{\num[round-precision=2]{+10.952}} \\ {\num[round-precision=1]{-8.654}}}\,\right)$ & {\num[round-precision=3]{170.850}} $\left(\,{\num[round-precision=3]{222.320}}\substack{{\num[round-precision=1]{+47.490}} \\ {\num[round-precision=2]{-32.980}}}\,\right)$ \\
125 & 62     & $ 0.490\pm0.004 $ & {\num[round-precision=2]{24.128}} $\left(\,{\num[round-precision=2]{27.429}}\substack{{\num[round-precision=1]{+8.359}} \\ {\num[round-precision=1]{-6.605}}}\,\right)$ & {\num[round-precision=3]{118.090}} $\left(\,{\num[round-precision=3]{154.830}}\substack{{\num[round-precision=2]{+22.810}} \\ {\num[round-precision=2]{-15.370}}}\,\right)$ \\
300 & 100    & $ 0.643\pm0.003 $ & {\num[round-precision=2]{18.387}} $\left(\,{\num[round-precision=2]{20.903}}\substack{{\num[round-precision=1]{+6.370}} \\ {\num[round-precision=1]{-5.033}}}\,\right)$ & {\num[round-precision=2]{28.950}} $\left(\,{\num[round-precision=2]{35.450}}\substack{{\num[round-precision=1]{+9.290}} \\ {\num[round-precision=1]{-6.500}}}\,\right)$ \\
600 & 100    & $ 0.688\pm0.003 $ & {\num[round-precision=2]{17.184}} $\left(\,{\num[round-precision=2]{19.535}}\substack{{\num[round-precision=1]{+5.954}} \\ {\num[round-precision=1]{-4.704}}}\,\right)$ & {\num[round-precision=2]{26.840}} $\left(\,{\num[round-precision=2]{33.650}}\substack{{\num[round-precision=1]{+7.430}} \\ {\num[round-precision=1]{-6.680}}}\,\right)$ \\
900 & 100    & $ 0.680\pm0.003 $ & {\num[round-precision=2]{17.386}} $\left(\,{\num[round-precision=2]{19.765}}\substack{{\num[round-precision=1]{+6.024}} \\ {\num[round-precision=1]{-4.759}}}\,\right)$ & {\num[round-precision=2]{26.890}} $\left(\,{\num[round-precision=2]{32.840}}\substack{{\num[round-precision=1]{+7.470}} \\ {\num[round-precision=1]{-5.820}}}\,\right)$ \\
\hline
\multicolumn{2}{l}{$Z' \rightarrow a+\gamma \rightarrow 3\gamma$} & \multicolumn{1}{c}{} & \multicolumn{1}{c}{} & \multicolumn{1}{c}{} \\
$m_{Z'}$ [GeV] & $m_{a}$ [GeV] &   &   & \\
\hline
100  & 40    & $ 0.438\pm0.009 $ & {\num[round-precision=2]{26.992}} $\left(\,{\num[round-precision=2]{30.686}}\substack{{\num[round-precision=1]{+9.352}} \\ {\num[round-precision=1]{-7.389}}}\,\right)$ & {\num[round-precision=3]{316.370}} $\left(\,{\num[round-precision=3]{386.690}}\substack{{\num[round-precision=2]{+97.540}} \\ {\num[round-precision=2]{-74.700}}}\,\right)$ \\
200  & 100   & $ 0.611\pm0.005 $ & {\num[round-precision=2]{19.350}} $\left(\,{\num[round-precision=2]{21.997}}\substack{{\num[round-precision=1]{+6.704}} \\ {\num[round-precision=1]{-5.297}}}\,\right)$ & {\num[round-precision=2]{53.110}} $\left(\,{\num[round-precision=2]{61.540}}\substack{{\num[round-precision=2]{+19.610}} \\ {\num[round-precision=2]{-15.640}}}\,\right)$ \\
400  & 100   & $ 0.649\pm0.004 $ & {\num[round-precision=2]{18.217}} $\left(\,{\num[round-precision=2]{20.709}}\substack{{\num[round-precision=1]{+6.311}} \\ {\num[round-precision=1]{-4.987}}}\,\right)$ & {\num[round-precision=2]{50.880}} $\left(\,{\num[round-precision=2]{62.800}}\substack{{\num[round-precision=2]{+13.870}} \\ {\num[round-precision=2]{-11.050}}}\,\right)$ \\
600  & 100   & $ 0.667\pm0.004 $ & {\num[round-precision=2]{17.725}} $\left(\,{\num[round-precision=2]{20.151}}\substack{{\num[round-precision=1]{+6.141}} \\ {\num[round-precision=1]{-4.852}}}\,\right)$ & {\num[round-precision=2]{39.470}} $\left(\,{\num[round-precision=2]{48.250}}\substack{{\num[round-precision=2]{+11.920}} \\ {\num[round-precision=1]{-8.780}}}\,\right)$ \\
1000 & 100   & $ 0.636\pm0.004 $ & {\num[round-precision=2]{18.589}} $\left(\,{\num[round-precision=2]{21.133}}\substack{{\num[round-precision=1]{+6.440}} \\ {\num[round-precision=1]{-5.089}}}\,\right)$ & {\num[round-precision=2]{38.420}} $\left(\,{\num[round-precision=2]{46.090}}\substack{{\num[round-precision=2]{+11.280}} \\ {\num[round-precision=1]{-9.480}}}\,\right)$ \\

      \hline \hline
    \end{tabular}
  \caption{Top row: Observed and expected model-independent upper limits on event yields for new physics processes for the inclusive signal region.  Also shown are the efficiencies for the fiducial kinematic region defined in the text for some example mass points for the signal scenarios explicitly considered here, and the corresponding observed and expected ($\pm 1 \sigma$) upper limits on the fiducial cross section within the acceptance.  Total statistical uncertainties are quoted for the fiducial efficiencies, and the uncertainties for the upper limits correspond to the uncertainties arising from the $\pm 1 \sigma$ upper limits calculated via hypothesis testing using the combination of statistical and systematic uncertainties.}
  \label{tab:ResultsFidSR}
  \end{center}
\end{table}

Using the same data-driven and simulation-based methodology restricted to the region 80 GeV $< m_{3\gamma} < 100$~GeV provides a test for the rare decay of the $Z$ boson to three photons.  The SM branching ratio for the process is predicted to be $\sim$5$\times$10$^{-10}$~\cite{Glover:1993nv}, but it has yet to be observed.  Table~\ref{tab:ResultsExpObs} (right) and Figure~\ref{fig:ResultsTTT} (right) summarise the observed counts as well as background expectations in this restricted region.  The data are consistent with the SM expectation: 244 events are observed and 233 $\pm$ 28 events are expected, while the signal expectation from simulation, for BR$(Z \rightarrow 3\gamma) = 10^{-5}$ (corresponding to the previous limit from LEP~\cite{Acciarri:1994gb}), is 418 $\pm$ 9 events.  Using the same hypothesis-testing and limit-setting procedure described above, but taking the signal expectation from the simulated sample described in Section~\ref{sec:simulation}, the observed (expected) limit, at the 95\% C.L., on the branching ratio of the $Z$ boson decay to three photons is found to be BR$(Z \rightarrow 3\gamma) <$ 2.2 (2.0) $\times 10^{-6}$, a result five times stronger than the previous LEP limit of $< 10^{-5}$.

\subsection{The $2\gamma$ and $3\gamma$ resonance searches} \label{sec:ressearch}

In addition to the tests based on the number of events in the inclusive signal regions, searches are performed for resonances in the two-photon and three-photon invariant mass ($m_{2\gamma}$ and $m_{3\gamma}$) distributions for events in the inclusive signal region. For these resonance searches, the background contribution is estimated from a fit to the $m_{2\gamma}$ or $m_{3\gamma}$ sideband regions, and thus does not rely upon simulated samples for the background estimate.  The sideband is modelled as a fourth-order polynomial, and the size of the sideband is mass-dependent, symmetric around the hypothesised resonance mass, following a local-spectrum approach.  The range of the observed mass spectrum that is used for the sideband fit is a local, truncated subset of the full spectrum.  For the $m_{2\gamma}$ ($m_{3\gamma}$) resonance search, the sideband is 20 (25) GeV in each direction for $m_{2\gamma}$ ($m_{3\gamma}$) $<$ 90 (230) GeV, where the event counts change rapidly as a function of $m_{2\gamma}$ ($m_{3\gamma}$), and rises to a sideband size of 80 (100) GeV in each direction for $m_{2\gamma}$ ($m_{3\gamma}$) $>$ 195 (425)~GeV, increasing roughly linearly with mass as the spectrum becomes smoother.  The $m_{2\gamma}$ ($m_{3\gamma}$) resonance search begins at a mass hypothesis of 10 (100) GeV, and proceeds in steps of 0.5 GeV.  The signal component of the resonance search is a Gaussian function with a fixed width that varies with particle mass, and the widths are determined from simulated signal samples.  Since the simulated signal samples are generated with a narrow-width approximation for both the pseudoscalar $a$ and the $Z'$ in all cases, the $2\gamma$ and $3\gamma$ mass resolutions for this search are equivalent to Gaussian functions that account for detector resolution, and are determined via fits to the simulated signal samples.  Hypothesis testing and limit setting are performed using the profile likelihood ratio as the test statistic for the $CL_{s}$ technique in the asymptotic approximation.

\begin{figure}
  \centering
  \subfigure{
  \includegraphics[width=0.45\columnwidth]{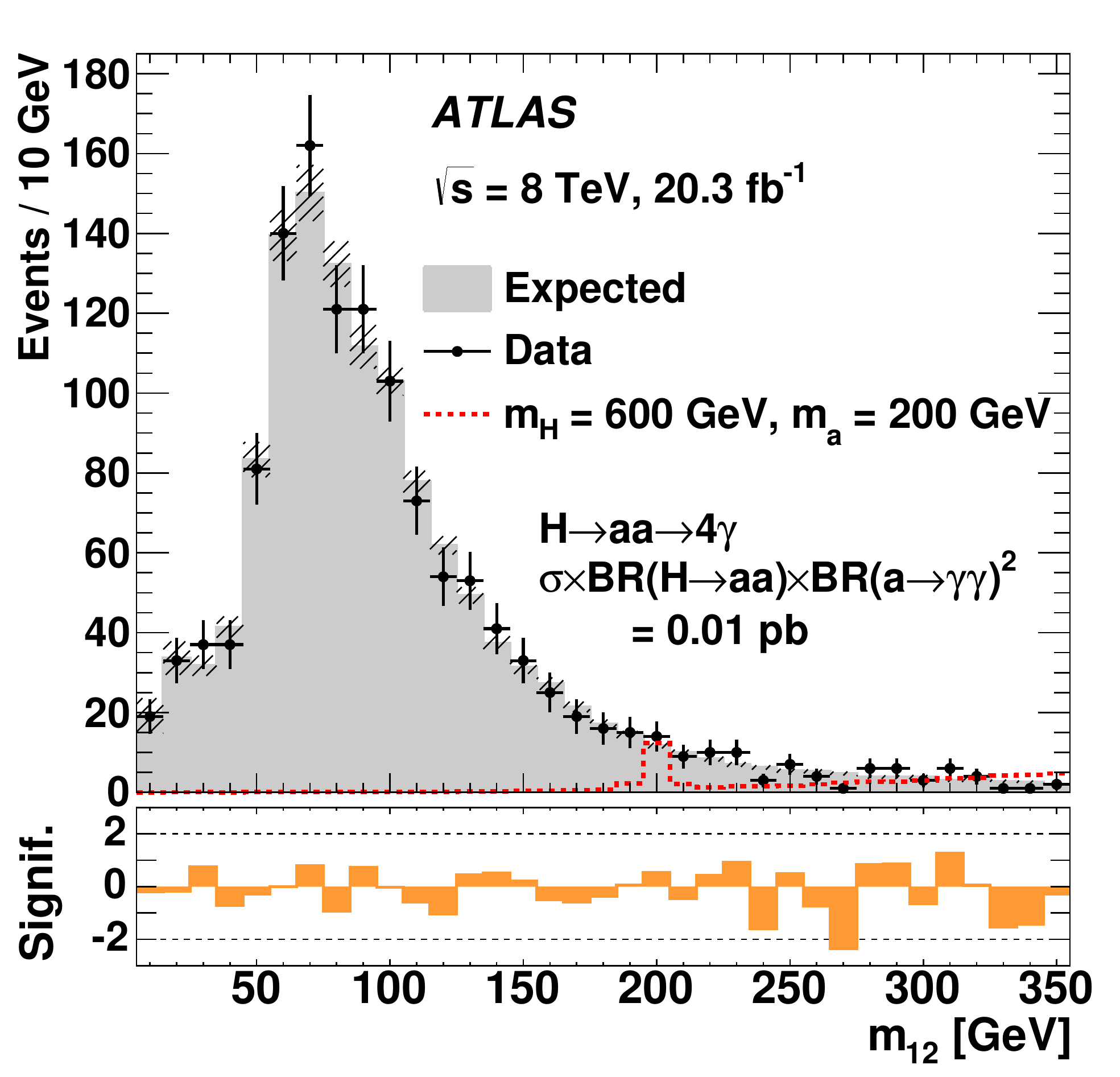}
  }
  \hskip 10mm
  \subfigure{
  \includegraphics[width=0.45\columnwidth]{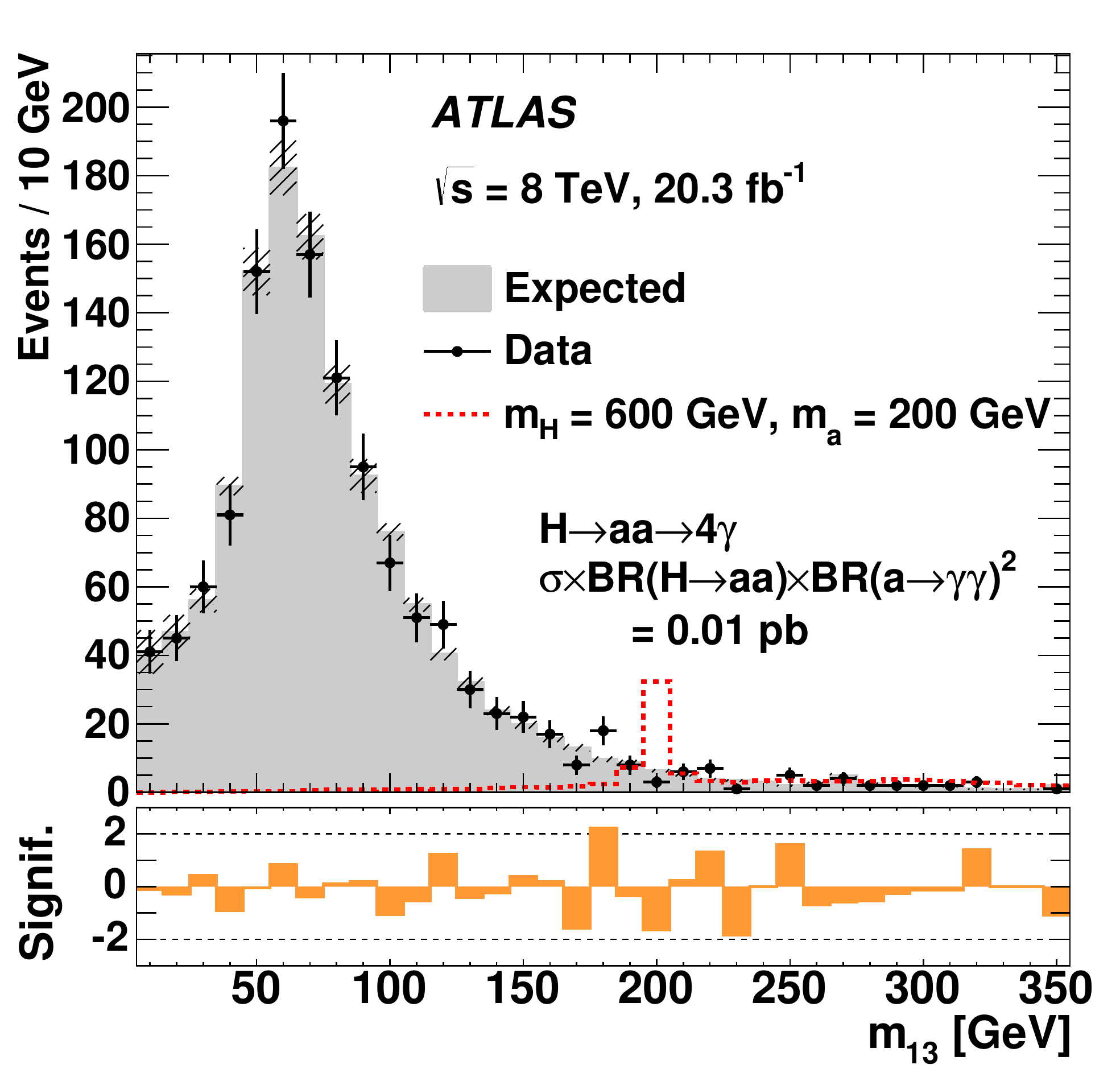}
  }
  \hskip 10mm
  \subfigure{
  \includegraphics[width=0.45\columnwidth]{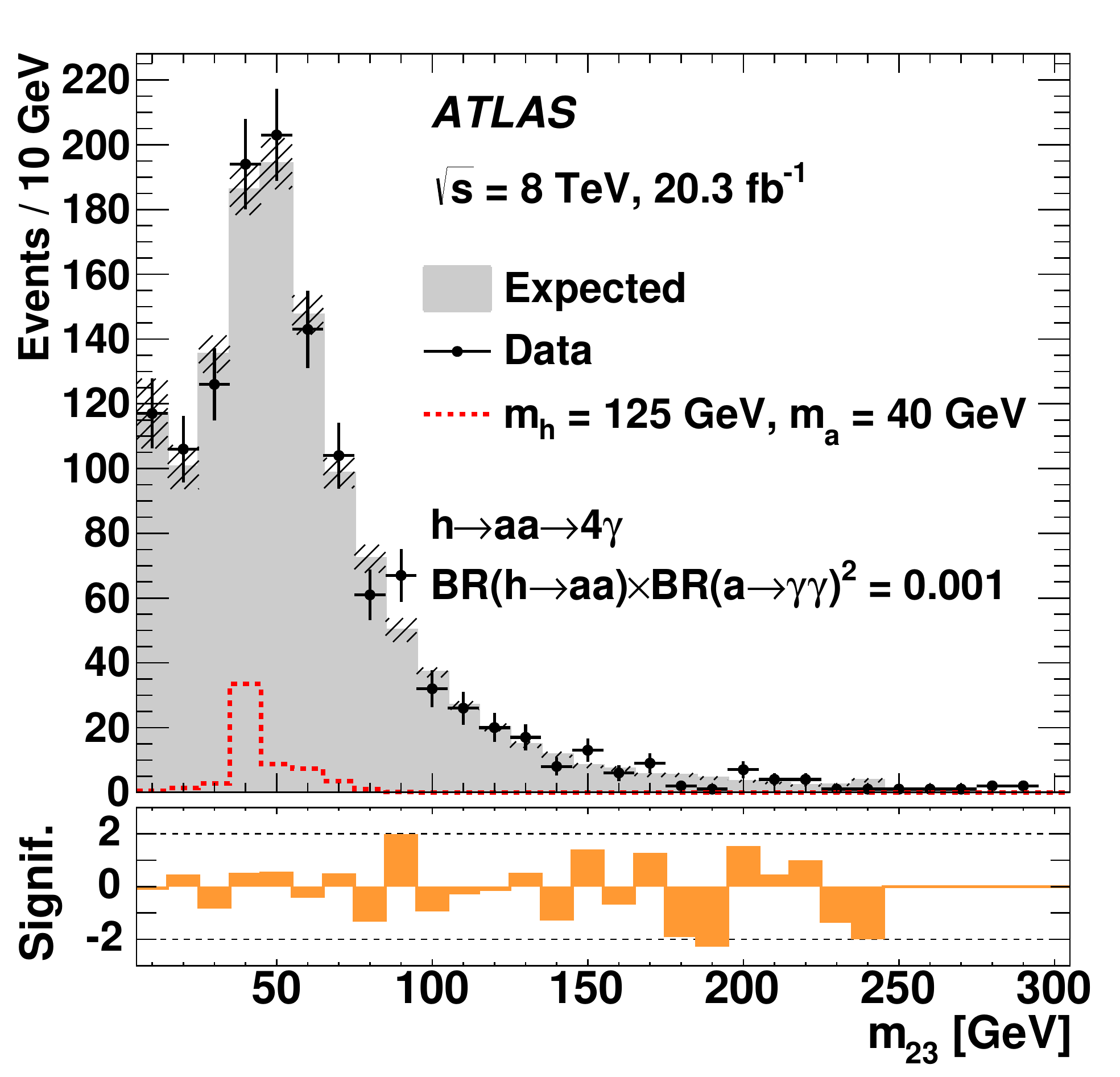}
  }
  \hskip 10mm
  \subfigure{
  \includegraphics[width=0.45\columnwidth]{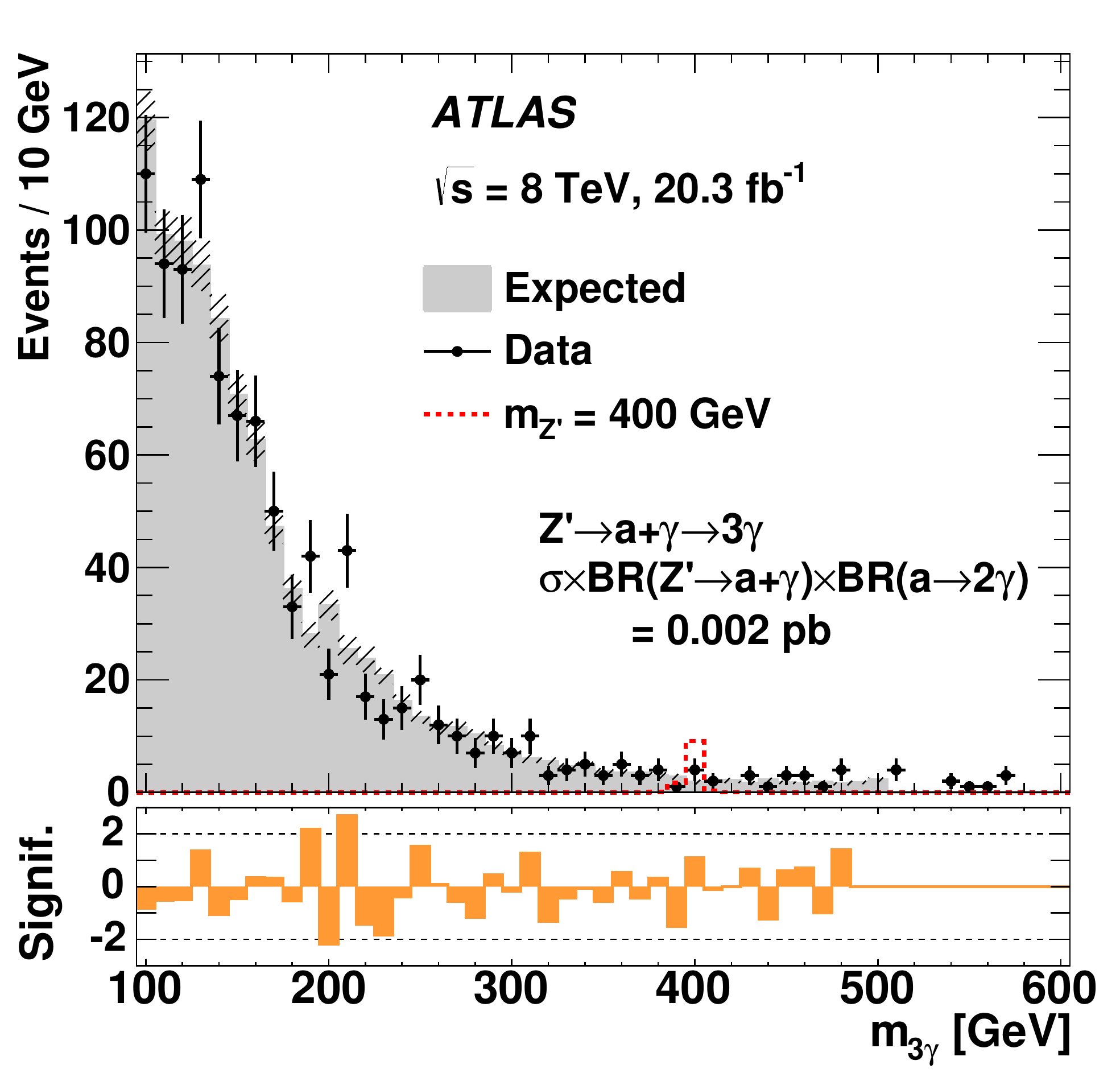}
  }
  \caption{Observed spectra of $m_{12}$, $m_{13}$, and $m_{23}$, where the 1, 2, and 3 refer to three \pT-ordered photons, as well as $m_{3\gamma}$.  For illustration purposes only, also shown is the expected background per bin, determined via unbinned sideband fits to the data as a part of the resonance search, for a hypothesised resonance mass defined by the centre of the bin, as well as the signal expectation for a few mass points for the BSM scenarios considered here.  The lower panels show the significance, in units of standard deviations of a Gaussian function, of the observation in each bin, taking into account the fractional uncertainty on the background as a result of the sideband fit.  This significance is derived from the $p$-value for the background-only hypothesis for each bin, calculated using a frequentist binomial parameter test~\cite{2008NIMPA.595..480C,2006sppp.conf..112C,2003sppp.conf...35L}.  The signal distributions used for the $m_{2\gamma}$ resonance searches have two components, a narrow Gaussian core for correctly paired two-photon combinations and a wide distribution for incorrectly paired combinations that is well described by the polynomial used to simultaneously model the background shape for the resonance search described in Section~\ref{sec:ressearch}.}
    \label{fig:ObsSRMassSpectra}
\end{figure}

\begin{figure}
  \centering
  \subfigure{
  \includegraphics[width=0.45\columnwidth]{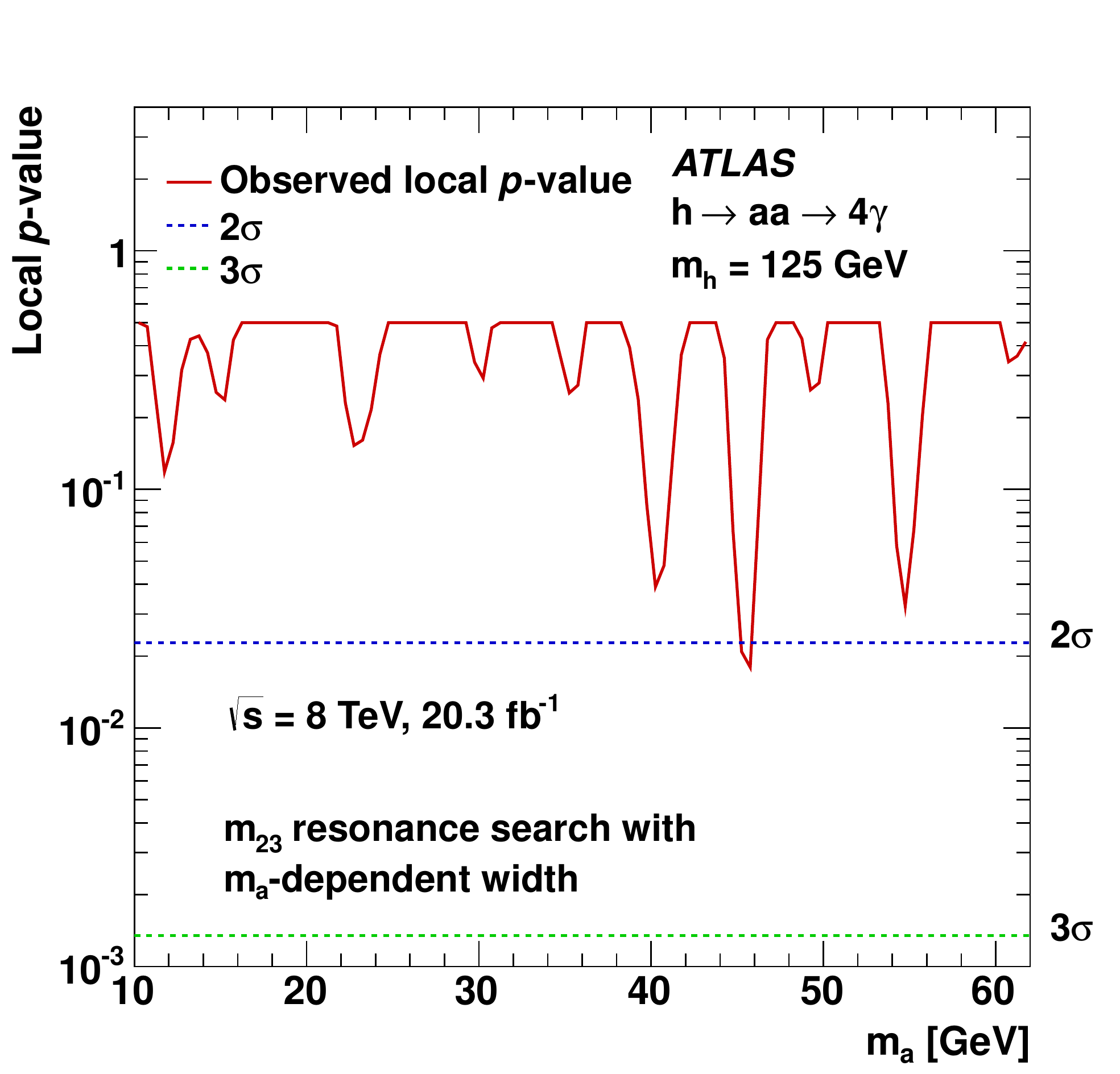}
  }
  \hskip 10mm
  \subfigure{
  \includegraphics[width=0.45\columnwidth]{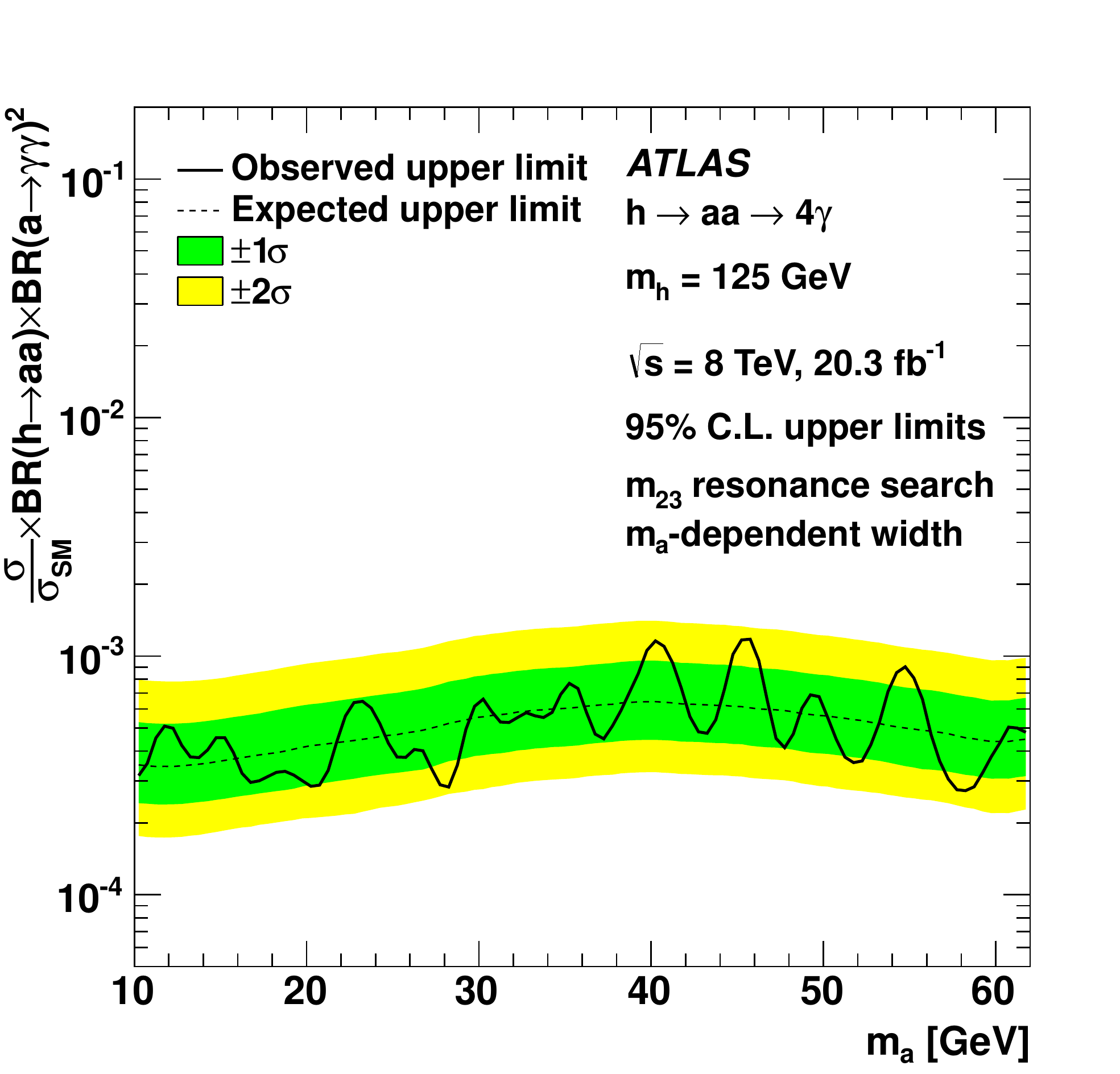}
  }
  \hskip 10mm
  \subfigure{
  \includegraphics[width=0.45\columnwidth]{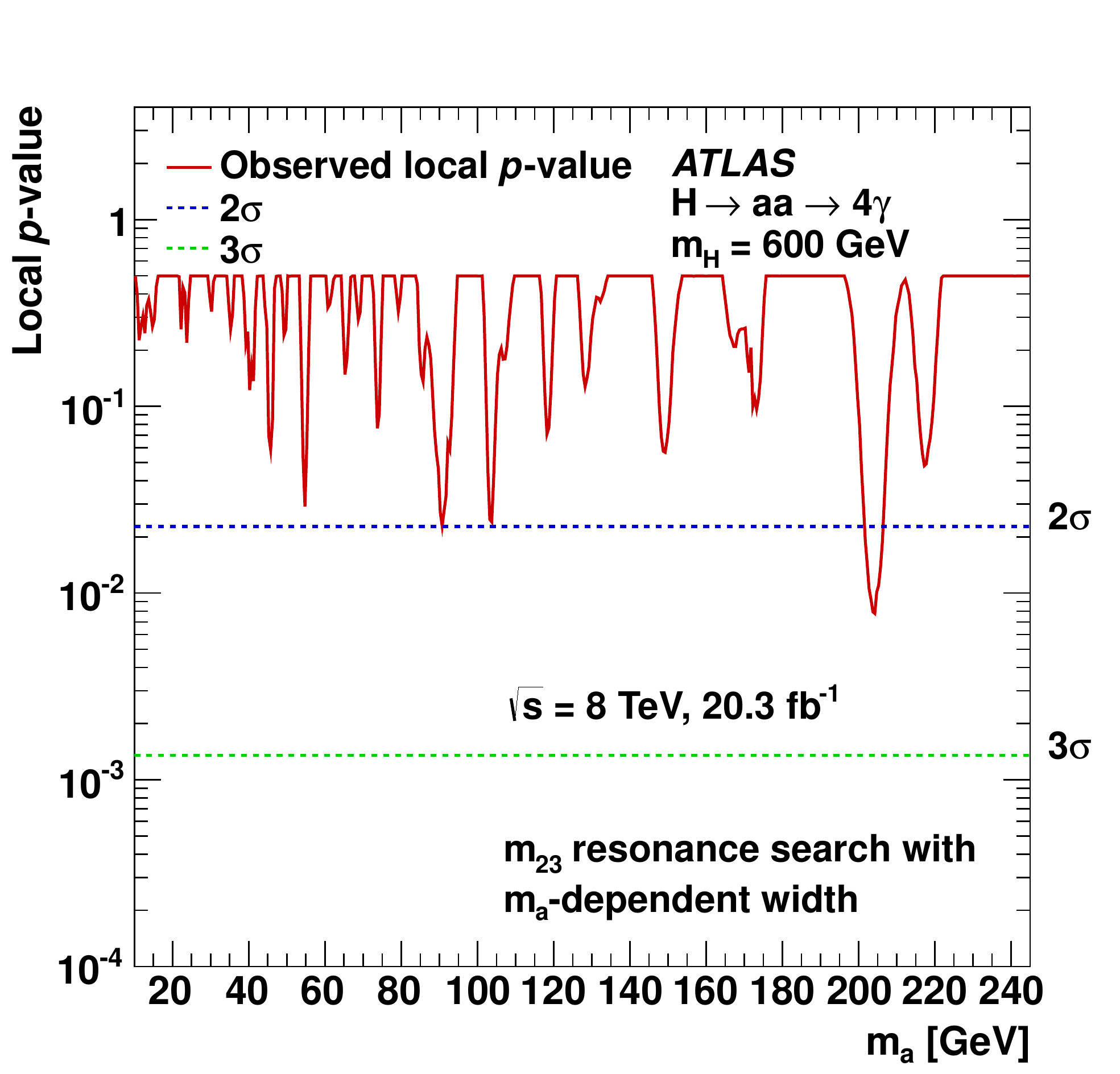}
  }
  \hskip 10mm
  \subfigure{
  \includegraphics[width=0.45\columnwidth]{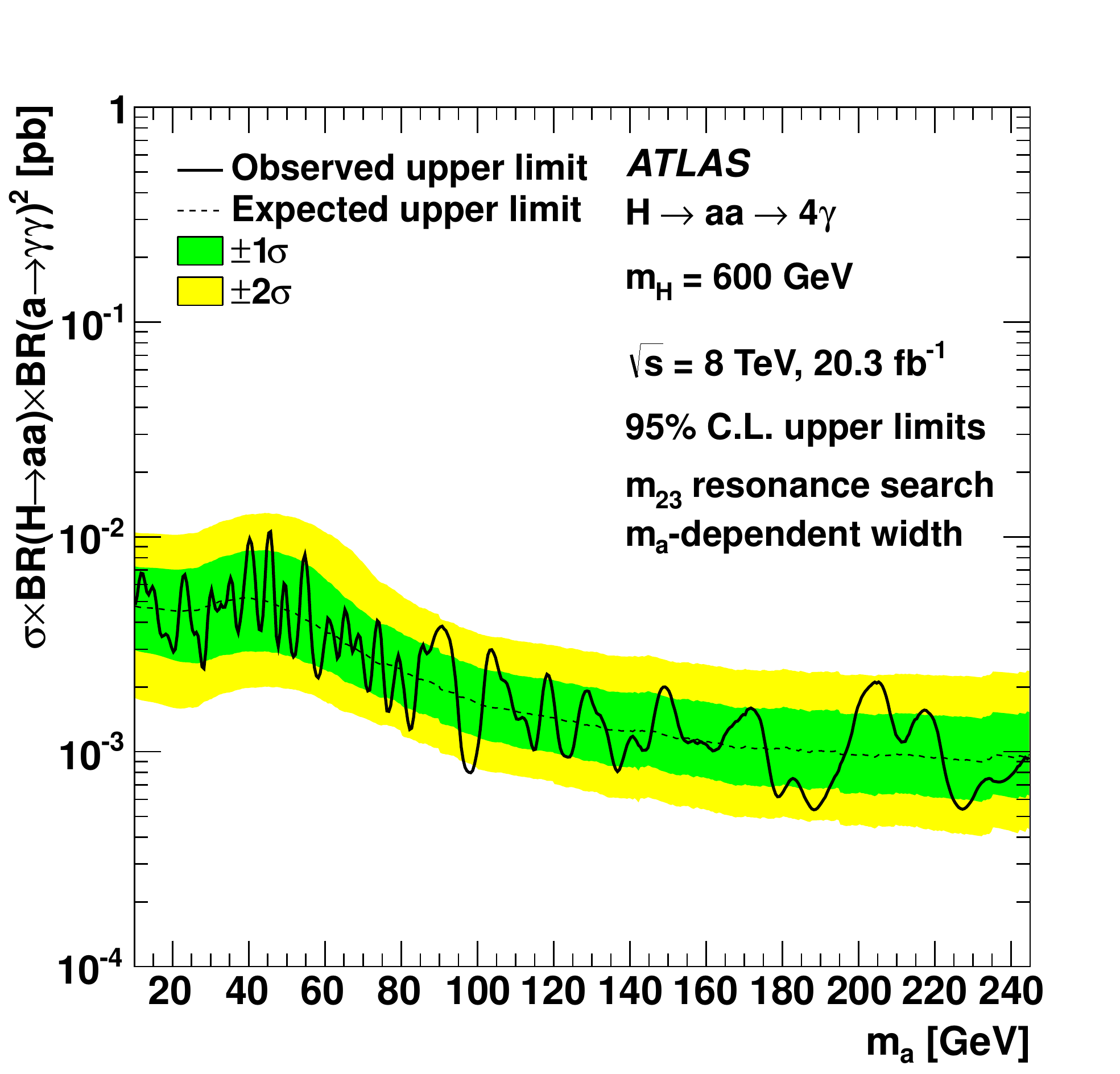}
  }
  \caption{Left: Local $p$-values for the background-only hypothesis as a result of a resonance search with respect to the BSM process $h/H \rightarrow aa \rightarrow 4\gamma$, for $m_{h}$ = 125 GeV (top row) and $m_{H}$ = 600 GeV (bottom row), as a function of $m_{a}$, determined via a search for local excesses in the $m_{23}$ spectrum.  Right: Upper limits, at the 95\% C.L., on $(\sigma / \sigma_{\text{SM}}) \times {\text {BR}}(h \rightarrow aa) \times {\text {BR}}(a \rightarrow \gamma\gamma)^{2}$ (top row) and $\sigma_{H} \times {\text {BR}}(H \rightarrow aa) \times {\text {BR}}(a \rightarrow \gamma\gamma)^{2}$ (bottom row).  Also shown are the $\pm 1$ and $2 \sigma$ uncertainty bands resulting from the resonance search hypothesis tests, taking into account the statistical and systematic uncertainties from simulated signal samples which are used to determine signal efficiency and Gaussian resonance width due to detector resolution for each mass hypothesis.}
    \label{fig:BSMHiggsPair23BRExpUL}
\end{figure}

\begin{figure}
  \centering
  \subfigure{
  \includegraphics[width=0.47\columnwidth]{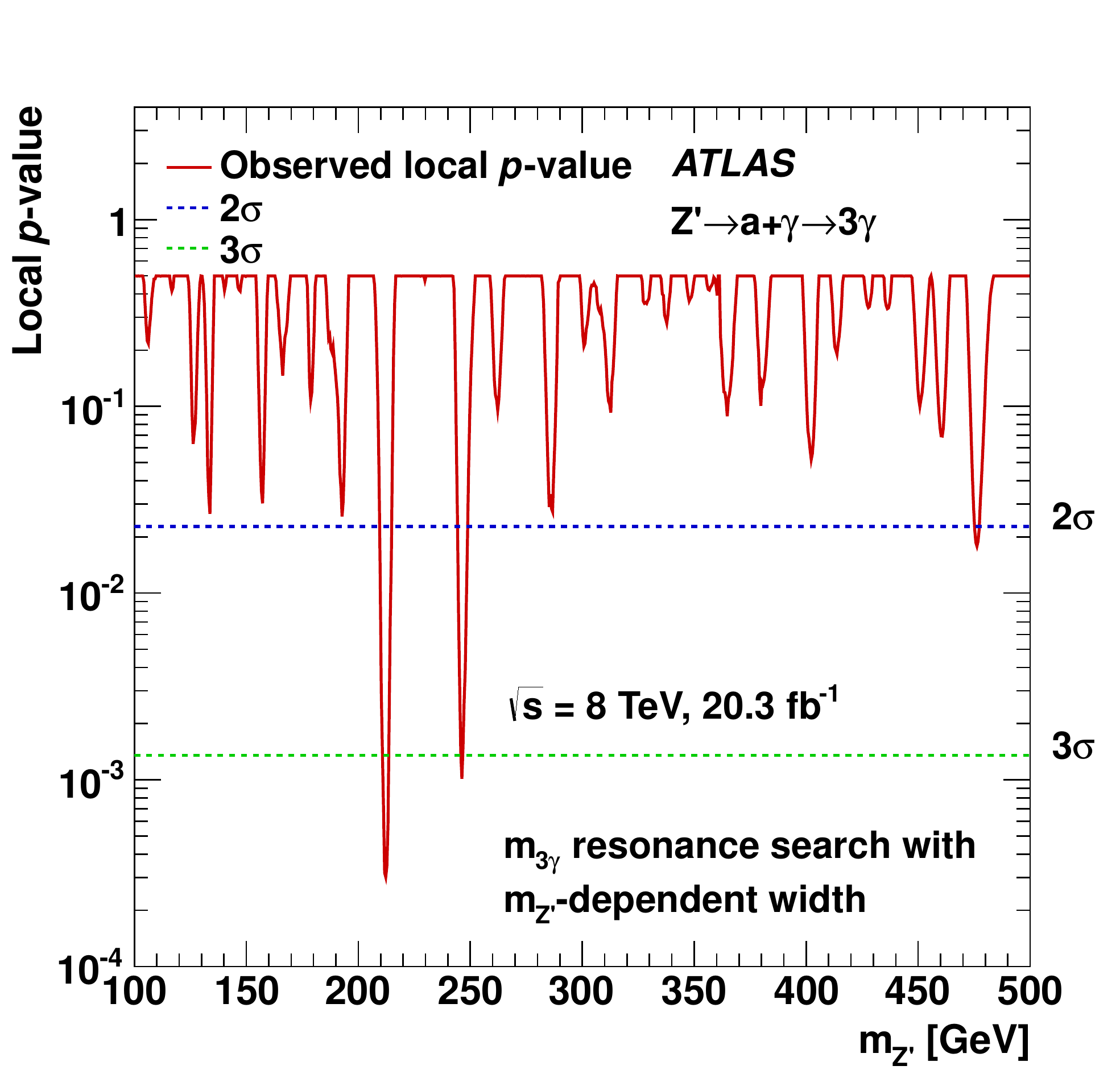}
  }
  \hskip 2mm
  \subfigure{
  \includegraphics[width=0.47\columnwidth]{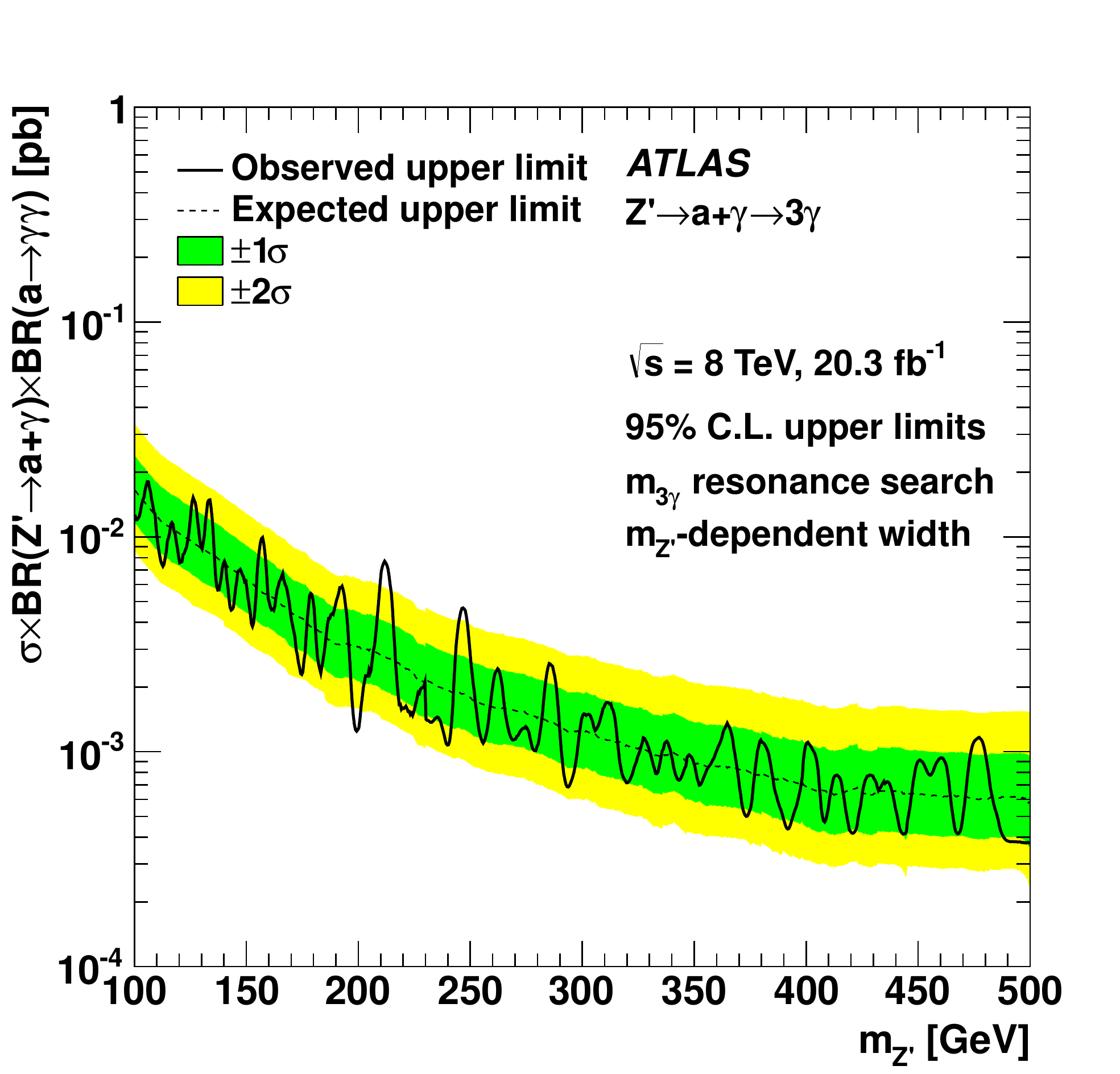}
  }
  \caption{Left: Local $p$-values for the background-only hypothesis as a result of a resonance search with respect to the production of a new vector gauge boson $Z'$ as a function of $m_{Z'}$, determined via a search for local excesses in the $m_{3\gamma}$ spectrum, using a narrow-width approximation to the $Z'$ resonance width.  The smallest local $p$-value is found to be 0.0003 ($3.4 \sigma$) which corresponds to a global $p$-value of 0.087 ($1.4 \sigma$).  Right: Upper limits, at the 95\% C.L., on $\sigma_{Z'} \times {\text {BR}}(Z' \rightarrow a+\gamma) \times {\text {BR}}(a \rightarrow \gamma\gamma)$.  Also shown are the $\pm 1$ and $2 \sigma$ uncertainty bands resulting from the resonance search hypothesis tests, taking into account the statistical and systematic uncertainties from simulated signal samples which are used to determine signal efficiency and Gaussian resonance width due to detector resolution for each mass hypothesis.}
  \label{fig:ZprimeXsecBRExpULResSearch}
\end{figure}

The resonance search is performed separately for the three two-photon mass spectra defined by the three possible photon pairings for three photons in the inclusive signal region.  As mentioned previously, the photons are ordered by \pT, from highest to lowest, and so the three two-photon mass spectra are denoted $m_{12}$, $m_{13}$ and $m_{23}$, where the 1, 2, and 3 refer to the \pT-ordered photons.  The observed $m_{2\gamma}$ and $m_{3\gamma}$ spectra in the inclusive signal region are shown in Figure~\ref{fig:ObsSRMassSpectra}.  Also shown in Figure~\ref{fig:ObsSRMassSpectra}, for visualisation purposes only, is the background expectation per bin, determined from the sideband fit to data as a part of the resonance search.  The resonance search is performed with a step size of 0.5 GeV and so the final results shown in Figures~\ref{fig:BSMHiggsPair23BRExpUL} and~\ref{fig:ZprimeXsecBRExpULResSearch} demonstrate sensitivity to resonances with widths appropriate to the BSM models considered here.  The widths of the bins in Figure~\ref{fig:ObsSRMassSpectra} do not correspond to the mass resolution for the signal scenarios in question.  The background estimates and significances shown in Figure~\ref{fig:ObsSRMassSpectra} provide a complementary comparison of the local agreement between data and expectation.  The lower panels show the significance, in units of standard deviations of a Gaussian function, of the observation in each bin, taking into account the fractional uncertainty on the background as a result of the sideband fit.  The significances shown in the lower panels in Figure~\ref{fig:ObsSRMassSpectra} are derived from the $p$-value for the background-only hypothesis for each bin, calculated using a frequentist binomial parameter test~\cite{2008NIMPA.595..480C,2006sppp.conf..112C,2003sppp.conf...35L}.  For regions beyond the sensitivity of the search, no background estimate is shown.

For the $H/h$ $\rightarrow aa \rightarrow 4\gamma$ BSM signal scenario, the photon pairing (among the three \pT-ordered photons) that most often corresponds to the photons arising from the decay of the same $a$ particle is (2,3).  As a result, the resonance search in the $m_{23}$ spectrum provides the best sensitivity to this model.  The widths of the Gaussian signal component --- corresponding to the detector resolution --- are taken from simulated samples of a Higgs boson decaying to four photons via a pair of intermediate pseudoscalar $a$ particles, and vary from 0.6 GeV $< \sigma_{\text{Gauss}} <$ 3.2 GeV for 10 GeV $< m_{a} <$ 440 GeV, and are largely independent of $m_{h/H}$.

No excess above background is detected, and upper limits, for a SM-like Higgs boson of $m_{h} = 125$ GeV (and assuming kinematics associated only with gluon fusion SM Higgs boson production), are calculated.  Additionally, limits are set for Higgs boson-like scalars with masses larger than 125 GeV.  The results of these resonance searches are shown in Figure~\ref{fig:BSMHiggsPair23BRExpUL} for the SM-like Higgs boson of $m_{h} = 125$ GeV (in the top row) and, as an example of a higher scalar mass, for $m_{H} = 600$ GeV (in the bottom row).  The resonance search limits for higher values of $m_{H}$ are limited by the small number of events in the mass spectra at higher values of $m_{2\gamma}$.   As shown in Figure~\ref{fig:BSMHiggsPair23BRExpUL}, the limits vary as a function of two-photon invariant mass, but an overall upper bound on the limits is determined to be $\sigma \times {\text {BR}}(h \rightarrow aa) \times {\text {BR}}(a \rightarrow \gamma \gamma)^{2}$ $< 1 \times 10^{-3} \sigma_{\text{SM}}$, for 10 GeV $< m_{a} <$ 62 GeV for the SM-like Higgs boson of $m_{h} = 125$ GeV and, for the higher scalar mass case, $\sigma_{H} \times {\text {BR}}(H \rightarrow aa) \times {\text {BR}}(a \rightarrow \gamma \gamma)^{2} <$ 0.02 pb for lower $m_{a}$ values in the range 10 GeV $< m_{a} <$ 90 GeV and $<$ 0.001 pb for higher $m_{a}$ (up to 245 GeV for the resonance search in the $m_{23}$ spectrum for $m_{H} = 600$ GeV shown in Figure~\ref{fig:BSMHiggsPair23BRExpUL}).  Additionally, using the expected signal yields from simulated samples, inclusive limits are calculated for 300 GeV $< m_{H} <$ 900 GeV and for a range of $m_{a} < m_{H}/2$, including values beyond the range of the mass spectra used for the resonance search.  These inclusive limits are shown in Table~\ref{tab:ResultsExpObsNCHH}.

Moreover, upper limits on the $Z'$ production cross section times the product of branching ratios, $\sigma_{Z'} \times {\text {BR}}(Z' \rightarrow a+\gamma) \times {\text {BR}}(a \rightarrow \gamma\gamma)$, are found to be in the range of 0.04 pb to 0.3 pb, depending upon $m_{Z'}$ and $m_{a}$.  Upper limits, at the 95\% C.L., on $\sigma_{Z'} \times {\text {BR}}(Z' \rightarrow a+\gamma) \times {\text {BR}}(a \rightarrow \gamma\gamma)$ are shown in Table~\ref{tab:ResultsExpObsNCZprime}, as a function of $m_{a}$, using the expected signal yields from simulated samples.  Additionally, using a narrow-width approximation to the $Z'$ resonance width, local excesses corresponding to Gaussian resonances due to detector resolution are searched for in the $m_{3\gamma}$ spectrum.  The Gaussian widths are determined via fits to the $Z'$ simulated signal samples, described in Section~\ref{sec:simulation}.  For the range of $m_{Z'}$ for which the resonance search is possible, the $Z'$ width exhibits a small dependence on $m_{a}$.  For each $Z'$ mass point, three different samples are simulated, with different values of $m_{a}$.  The average of the three measured $Z'$ widths for each of the $m_{a}$ points simulated is taken as the width for a given $m_{Z'}$, and these values are used for the three-photon resonance search, interpolating for $m_{Z'}$ points between those for which samples are simulated.  These values range from 1.5 GeV $< \sigma_{\text{Gauss}} <$ 2.4 GeV for 100 GeV $< m_{Z'} <$ 500 GeV.  The results, along with the local $p$-values for the background-only hypothesis, are shown in Figure~\ref{fig:ZprimeXsecBRExpULResSearch}.  The smallest local $p$-value is found to be 0.0003 ($3.4 \sigma$ local significance), at $m_{3\gamma} =$ 212 GeV which, after adjusting for a trials factor~\cite{Gross:2010qma}, corresponds to a global $p$-value of 0.087 ($1.4 \sigma$ global significance).

\begin{table}
  \begin{center}
  \sisetup{retain-explicit-plus}
  \sisetup{round-mode = figures,round-precision=2,retain-explicit-plus=true}
    \begin{tabular}{ l c c c c }
      \hline \hline
      \multicolumn{5}{c}{Observed (expected) 95\% C.L. upper limits } \\ 
	\hline
      \multicolumn{1}{c}{ $m_{H}$ [GeV] } & \multicolumn{4}{c}{$m_{a}$ [GeV]} \\
      \multicolumn{1}{c}{} & \multicolumn{4}{c}{$\sigma_{H} \times {\text {BR}}(H \rightarrow aa) \times {\text {BR}}(a \rightarrow \gamma\gamma)^{2}$ [fb]} \\
      	\hline
     & 20 & 50 & 100 & 140 \\
 300 & {\num[round-precision=2]{48.100}} $\left(\,{\num[round-precision=2]{60.370}}\substack{{\num[round-precision=2]{+12.960}} \\ {\num[round-precision=1]{-10.370}}}\,\right)$ & {\num[round-precision=2]{33.310}} $\left(\,{\num[round-precision=2]{40.370}}\substack{{\num[round-precision=1]{+9.480}} \\ {\num[round-precision=1]{-7.950}}}\,\right)$ & {\num[round-precision=2]{28.950}} $\left(\,{\num[round-precision=2]{35.450}}\substack{{\num[round-precision=1]{+9.290}} \\ {\num[round-precision=1]{-6.500}}}\,\right)$ & {\num[round-precision=2]{27.740}} $\left(\,{\num[round-precision=2]{34.060}}\substack{{\num[round-precision=1]{+7.520}} \\ {\num[round-precision=1]{-6.320}}}\,\right)$ \\
 \hline 
     & 50 & 100 & 200 & 290 \\
 600 & {\num[round-precision=2]{31.050}} $\left(\,{\num[round-precision=2]{37.870}}\substack{{\num[round-precision=1]{+10.020}} \\ {\num[round-precision=1]{-6.820}}}\,\right)$ & {\num[round-precision=2]{26.840}} $\left(\,{\num[round-precision=2]{33.650}}\substack{{\num[round-precision=1]{+7.430}} \\ {\num[round-precision=1]{-6.680}}}\,\right)$ & {\num[round-precision=2]{25.230}} $\left(\,{\num[round-precision=2]{31.050}}\substack{{\num[round-precision=1]{+6.820}} \\ {\num[round-precision=1]{-5.700}}}\,\right)$ & {\num[round-precision=2]{25.230}} $\left(\,{\num[round-precision=2]{31.050}}\substack{{\num[round-precision=1]{+6.820}} \\ {\num[round-precision=1]{-5.700}}}\,\right)$ \\
 \hline 
     & 50 & 100 & 200 & 440 \\
 900 & {\num[round-precision=2]{35.900}} $\left(\,{\num[round-precision=2]{44.240}}\substack{{\num[round-precision=2]{+11.360}} \\ {\num[round-precision=1]{-8.340}}}\,\right)$ & {\num[round-precision=2]{26.890}} $\left(\,{\num[round-precision=2]{32.840}}\substack{{\num[round-precision=1]{+7.470}} \\ {\num[round-precision=1]{-5.820}}}\,\right)$ & {\num[round-precision=2]{26.420}} $\left(\,{\num[round-precision=2]{32.840}}\substack{{\num[round-precision=1]{+6.520}} \\ {\num[round-precision=1]{-5.950}}}\,\right)$ & {\num[round-precision=2]{25.530}} $\left(\,{\num[round-precision=2]{31.790}}\substack{{\num[round-precision=1]{+6.890}} \\ {\num[round-precision=1]{-5.260}}}\,\right)$ \\
 \hline

      \hline \hline
    \end{tabular}
  \caption{Expected and observed 95\% C.L. upper limits on $\sigma_{H} \times {\text {BR}}(H \rightarrow aa) \times {\text {BR}}(a \rightarrow \gamma\gamma)^{2}$.  The uncertainties for the expected limits are the $\pm 1 \sigma$ uncertainties resulting from the hypothesis tests for each mass point, taking into account statistical and systematic uncertainties.}
  \label{tab:ResultsExpObsNCHH}
  \end{center}
\end{table}

\begin{table}
  \begin{center}
  \sisetup{retain-explicit-plus}
  \sisetup{round-mode = figures,round-precision=2,retain-explicit-plus=true}
    \begin{tabular}{ l c c c }
      \hline \hline
      \multicolumn{4}{c}{Observed (expected) 95\% C.L. upper limits } \\
	\hline
      \multicolumn{1}{c}{ $m_{Z'}$ [GeV] } & \multicolumn{3}{c}{$m_{a}$ [GeV]} \\
      \multicolumn{1}{c}{} & \multicolumn{3}{c}{$\sigma_{Z'} \times {\text {BR}}(Z' \rightarrow a+\gamma) \times {\text {BR}}(a \rightarrow \gamma\gamma)$ [fb]} \\
      	\hline
	
     & 40 & 60 & 80 \\
 100 & {\num[round-precision=2]{316.370}} $\left(\,{\num[round-precision=2]{386.690}}\substack{{\num[round-precision=2]{+97.540}} \\ {\num[round-precision=1]{-74.700}}}\,\right)$ & {\num[round-precision=2]{148.860}} $\left(\,{\num[round-precision=2]{172.510}}\substack{{\num[round-precision=1]{+53.460}} \\ {\num[round-precision=1]{-39.380}}}\,\right)$ & {\num[round-precision=2]{313.640}} $\left(\,{\num[round-precision=2]{373.040}}\substack{{\num[round-precision=2]{+104.750}} \\ {\num[round-precision=1]{-80.190}}}\,\right)$ \\
 \hline 
     & 50 & 100 & 150 \\
 200 & {\num[round-precision=2]{78.320}} $\left(\,{\num[round-precision=2]{89.790}}\substack{{\num[round-precision=2]{+27.610}} \\ {\num[round-precision=2]{-21.950}}}\,\right)$ & {\num[round-precision=2]{53.110}} $\left(\,{\num[round-precision=2]{61.540}}\substack{{\num[round-precision=2]{+19.610}} \\ {\num[round-precision=2]{-15.640}}}\,\right)$ & {\num[round-precision=2]{51.120}} $\left(\,{\num[round-precision=2]{58.320}}\substack{{\num[round-precision=2]{+18.850}} \\ {\num[round-precision=2]{-13.650}}}\,\right)$ \\
 \hline 
     & 100 & 200 & 300 \\
 400 & {\num[round-precision=2]{50.880}} $\left(\,{\num[round-precision=2]{62.800}}\substack{{\num[round-precision=2]{+13.870}} \\ {\num[round-precision=1]{-11.050}}}\,\right)$ & {\num[round-precision=2]{44.260}} $\left(\,{\num[round-precision=2]{54.800}}\substack{{\num[round-precision=2]{+11.520}} \\ {\num[round-precision=1]{-8.750}}}\,\right)$ & {\num[round-precision=2]{38.470}} $\left(\,{\num[round-precision=2]{47.160}}\substack{{\num[round-precision=2]{+11.580}} \\ {\num[round-precision=1]{-9.790}}}\,\right)$ \\
 \hline 
     & 100 & 250 & 400 \\
 600 & {\num[round-precision=2]{39.470}} $\left(\,{\num[round-precision=2]{48.250}}\substack{{\num[round-precision=2]{+11.920}} \\ {\num[round-precision=1]{-8.780}}}\,\right)$ & {\num[round-precision=2]{41.050}} $\left(\,{\num[round-precision=2]{51.580}}\substack{{\num[round-precision=1]{+9.520}} \\ {\num[round-precision=1]{-7.900}}}\,\right)$ & {\num[round-precision=2]{41.050}} $\left(\,{\num[round-precision=2]{51.580}}\substack{{\num[round-precision=2]{+10.530}} \\ {\num[round-precision=1]{-7.900}}}\,\right)$ \\
 \hline 
     & 100 & 350 & 600 \\
 800 & {\num[round-precision=2]{38.420}} $\left(\,{\num[round-precision=2]{46.090}}\substack{{\num[round-precision=2]{+11.280}} \\ {\num[round-precision=1]{-9.480}}}\,\right)$ & {\num[round-precision=2]{35.490}} $\left(\,{\num[round-precision=2]{43.160}}\substack{{\num[round-precision=1]{+9.470}} \\ {\num[round-precision=1]{-7.670}}}\,\right)$ & {\num[round-precision=2]{35.490}} $\left(\,{\num[round-precision=2]{43.160}}\substack{{\num[round-precision=1]{+9.470}} \\ {\num[round-precision=1]{-9.280}}}\,\right)$ \\
 \hline 
     & 100 & 450 & 800 \\
 1000 & {\num[round-precision=2]{38.420}} $\left(\,{\num[round-precision=2]{46.090}}\substack{{\num[round-precision=2]{+11.280}} \\ {\num[round-precision=1]{-9.480}}}\,\right)$ & {\num[round-precision=2]{54.440}} $\left(\,{\num[round-precision=2]{63.910}}\substack{{\num[round-precision=2]{+17.140}} \\ {\num[round-precision=1]{-13.090}}}\,\right)$ & {\num[round-precision=2]{36.610}} $\left(\,{\num[round-precision=2]{43.160}}\substack{{\num[round-precision=2]{+12.400}} \\ {\num[round-precision=1]{-7.670}}}\,\right)$ \\
 \hline

      \hline \hline
    \end{tabular}
  \caption{Expected and observed 95\% C.L. upper limits on $\sigma_{Z'} \times {\text {BR}}(Z' \rightarrow a+\gamma) \times {\text {BR}}(a \rightarrow \gamma\gamma)$.  The uncertainties for the expected limits are the $\pm 1 \sigma$ uncertainties resulting from the hypothesis tests for each mass point, taking into account statistical and systematic uncertainties.}
  \label{tab:ResultsExpObsNCZprime}
  \end{center}
\end{table}

\newpage
\clearpage
\section{Conclusion}
\label{sec:conclusion}

A search for new phenomena in events with at least three photons has been performed using 20.3 fb$^{-1}$ of LHC $pp$ collision data at $\sqrt{s}$ = 8 TeV collected with the ATLAS detector at CERN.  The SM background expectation is in agreement with the data, and is determined to be 1370 $\pm$ 140 events while 1290 events are observed.  The model-independent observed (expected) 95\% C.L. upper limit on the number of signal events is found to be 240 (273$^{+83}_{-66}$).  Upper limits at the 95\% C.L. are calculated on the fiducial cross section $\sigma_{\text{fid}}$ for events from non-SM processes for several signal scenarios.  The observed (expected) limit on the branching ratio of the $Z$ boson decay to three photons is found to be BR$(Z \rightarrow 3\gamma) <$ 2.2 (2.0) $\times 10^{-6}$, a result five times stronger than the previous result from LEP.

In addition, a search for local excesses in the two-photon and three-photon invariant mass distributions is conducted.  For the two-photon mass spectra, no significant excesses are detected, and the 95\% C.L. upper limit on $(\sigma / \sigma_{\text{SM}}) \times {\text {BR}}(h \rightarrow aa) \times {\text {BR}}(a \rightarrow \gamma \gamma)^{2}$ (assuming kinematics associated only with gluon fusion SM Higgs boson production) is calculated to vary from $\sim$3$\times$10$^{-4}$ to $\sim$4$\times$10$^{-4}$ for 10 GeV $< m_{a} <$ 62 GeV for a SM-like Higgs boson with a mass of $m_{h} =$ 125 GeV.  Limits are set for Higgs boson-like scalars $H$ with masses up to $m_{H} = 900$ GeV and are found to be $\sigma_{H} \times {\text {BR}}(H \rightarrow aa) \times {\text {BR}}(a \rightarrow \gamma \gamma)^{2} <$ 0.02 to 0.001 pb, depending upon $m_{H}$ and $m_{a}$.  For the three-photon mass spectrum, the resonance search is conducted in the context of a $Z'$ decaying to three photons.  The smallest local $p$-value is found to be 0.0003 ($3.4 \sigma$ local significance), at $m_{Z'} =$ 212 GeV which, after adjusting for a trials factor, corresponds to a global $p$-value of 0.09 ($1.4 \sigma$ global significance).  Upper limits at the 95\% C.L. on the $Z'$ production cross section times the product of branching ratios, $\sigma_{Z'} \times {\text {BR}}(Z' \rightarrow a+\gamma) \times {\text {BR}}(a \rightarrow \gamma\gamma)$, are found to be in the range of 0.04 pb to 0.3 pb, depending upon $m_{Z'}$.

These model-independent results are the first of their kind, as are the interpretations for a Higgs boson decaying to four photons via two intermediate pseudoscalar $a$ particles (for a SM-like Higgs boson of $m_{h} =$~125~GeV and for Higgs-like scalars of higher masses) and for three-photon resonances corresponding to a new vector gauge boson.

\section*{Acknowledgements}


We thank CERN for the very successful operation of the LHC, as well as the
support staff from our institutions without whom ATLAS could not be
operated efficiently.

We acknowledge the support of ANPCyT, Argentina; YerPhI, Armenia; ARC,
Australia; BMWFW and FWF, Austria; ANAS, Azerbaijan; SSTC, Belarus; CNPq and FAPESP,
Brazil; NSERC, NRC and CFI, Canada; CERN; CONICYT, Chile; CAS, MOST and NSFC,
China; COLCIENCIAS, Colombia; MSMT CR, MPO CR and VSC CR, Czech Republic;
DNRF, DNSRC and Lundbeck Foundation, Denmark; EPLANET, ERC and NSRF, European Union;
IN2P3-CNRS, CEA-DSM/IRFU, France; GNSF, Georgia; BMBF, DFG, HGF, MPG and AvH
Foundation, Germany; GSRT and NSRF, Greece; RGC, Hong Kong SAR, China; ISF, MINERVA, GIF, I-CORE and Benoziyo Center, Israel; INFN, Italy; MEXT and JSPS, Japan; CNRST, Morocco; FOM and NWO, Netherlands; BRF and RCN, Norway; MNiSW and NCN, Poland; GRICES and FCT, Portugal; MNE/IFA, Romania; MES of Russia and NRC KI, Russian Federation; JINR; MSTD,
Serbia; MSSR, Slovakia; ARRS and MIZ\v{S}, Slovenia; DST/NRF, South Africa;
MINECO, Spain; SRC and Wallenberg Foundation, Sweden; SER, SNSF and Cantons of
Bern and Geneva, Switzerland; NSC, Taiwan; TAEK, Turkey; STFC, the Royal
Society and Leverhulme Trust, United Kingdom; DOE and NSF, United States of
America.

The crucial computing support from all WLCG partners is acknowledged
gratefully, in particular from CERN and the ATLAS Tier-1 facilities at
TRIUMF (Canada), NDGF (Denmark, Norway, Sweden), CC-IN2P3 (France),
KIT/GridKA (Germany), INFN-CNAF (Italy), NL-T1 (Netherlands), PIC (Spain),
ASGC (Taiwan), RAL (UK) and BNL (USA) and in the Tier-2 facilities
worldwide.

\bibliographystyle{bibtex/bst/atlasBibStyleWithTitle}
\bibliography{MultiPhotonPaper}

\newpage
\begin{flushleft}
{\Large The ATLAS Collaboration}

\bigskip

G.~Aad$^{\rm 85}$,
B.~Abbott$^{\rm 113}$,
J.~Abdallah$^{\rm 151}$,
O.~Abdinov$^{\rm 11}$,
R.~Aben$^{\rm 107}$,
M.~Abolins$^{\rm 90}$,
O.S.~AbouZeid$^{\rm 158}$,
H.~Abramowicz$^{\rm 153}$,
H.~Abreu$^{\rm 152}$,
R.~Abreu$^{\rm 116}$
% ...
\bigskip
\\
% ...
$^{11}$ Institute of Physics, Azerbaijan Academy of Sciences, Baku, Azerbaijan\\
% ...
$^{85}$ CPPM, Aix-Marseille Universit{\'e} and CNRS/IN2P3, Marseille, France\\
% ...
$^{90}$ Department of Physics and Astronomy, Michigan State University, East Lansing MI, United States of America\\
% ...
$^{107}$ Nikhef National Institute for Subatomic Physics and University of Amsterdam, Amsterdam, Netherlands\\
% ...
$^{113}$ Homer L. Dodge Department of Physics and Astronomy, University of Oklahoma, Norman OK, United States of America\\
% ...
$^{116}$ Center for High Energy Physics, University of Oregon, Eugene OR, United States of America\\
% ...
$^{151}$ Institute of Physics, Academia Sinica, Taipei, Taiwan\\
$^{152}$ Department of Physics, Technion: Israel Institute of Technology, Haifa, Israel\\
$^{153}$ Raymond and Beverly Sackler School of Physics and Astronomy, Tel Aviv University, Tel Aviv, Israel\\
% ...
$^{158}$ Department of Physics, University of Toronto, Toronto ON, Canada\\
% ...
\end{flushleft}


\end{document}